\documentclass[acmsmall]{acmart}

\usepackage{graphicx}
\usepackage{caption}
\usepackage{subcaption}
\usepackage{hyperref}
\usepackage{adjustbox}
\usepackage{multirow}
\usepackage{pgfplots}
\usepackage{array}
\def\approxprop{%
  \def\p{%
    \setbox0=\vbox{\hbox{$\propto$}}%
    \ht0=0.6ex \box0 }%
  \def\s{%
    \vbox{\hbox{$\sim$}}%
  }%
  \mathrel{\raisebox{0.7ex}{%
      \mbox{$\underset{\s}{\p}$}%
    }}%
 }
\usepackage{hyperref}
\usepackage{algorithm}
\usepackage{tabularx} 
\usepackage[noend]{algpseudocode}
\usepackage{xcolor}
\PassOptionsToPackage{dvipsnames}{xcolor}
\usepackage{threeparttable}
\pgfplotsset{compat=1.3}
\usepackage{longtable} 
\usepackage{lscape}

\AtBeginDocument{%
  \providecommand\BibTeX{{%
    \normalfont B\kern-0.5em{\scshape i\kern-0.25em b}\kern-0.8em\TeX}}}

\setcopyright{acmcopyright}
\copyrightyear{2022}
\acmYear{2022}
\acmDOI{XXXXXXX.XXXXXXX}

\acmJournal{JACM}
\acmVolume{37}
\acmNumber{4}
\acmArticle{111}
\acmMonth{8}

\begin{document}

\title{Design and Analysis of High Performance Heterogeneous Block-based Approximate Adders}

\author{Ebrahim Farahmand}
\email{ebrahim.farahmand@eng.uk.ac.ir}
\author{Ali Mahani}
\email{amahani@uk.ac.ir}
\affiliation{%
  \institution{Department of Electrical Engineering, Shahid Bahonar University of Kerman}
  \state{Kerman}
  \country{Iran}
}

\author{Muhammad Abdullah Hanif}
\email{mh6117@nyu.edu}
\author{Muhammad Shafique}
\email{muhammad.shafique@nyu.edu}
\affiliation{%
  \institution{eBrain Lab, Division of Engineering, New York University Abu Dhabi}
  \state{Abu Dhabi}
  \country{UAE}
}

\renewcommand{\shortauthors}{Ebrahim, et al.}

\begin{abstract}
Approximate computing is an emerging paradigm to improve the power and performance efficiency of error-resilient applications. 
As adders are one of the key components in almost all processing systems, a significant amount of research has been carried out towards designing approximate adders that can offer better efficiency than conventional designs, however, at the cost of some accuracy loss. 
In this paper, we highlight a new class of energy-efficient approximate adders, namely Heterogeneous Block-based Approximate Adders (HBAA), and propose a generic configurable adder model that can be configured to represent a particular HBAA configuration. 
An HBAA, in general, is composed of heterogeneous sub-adder blocks of equal length, where each sub-adder can be an approximate sub-adder and have a different configuration. 
The sub-adders are mainly approximated through inexact logic and carry truncation. 
Compared to the existing design space, HBAAs provide additional design points that fall on the Pareto-front and offer a better quality-efficiency trade-off in certain scenarios. 
Furthermore, to enable efficient design space exploration based on user-defined constraints, we propose an analytical model to efficiently evaluate the Probability Mass Function (PMF) of approximation error and other error metrics, such as Mean Error Distance
(MED), Normalized Mean Error
Distance (NMED) and Error Rate (ER) of HBAAs. 
The results show that HBAA configurations can provide around $15\%$ reduction in area and up to $17\%$ reduction in energy compared to state-of-the-art approximate adders.
\end{abstract}



\maketitle

\section{Introduction}
\label{sec:intro} 
Nowadays, due to the high computational requirements of advanced applications, computing systems are becoming more-and-more resource hungry. 
Moreover, because of the energy/power, area, and cost requirement issues, most of the emerging applications cannot be deployed on resource-constrained edge devices. 
Approximate computing has achieved notable attention due to its potential to increase computing efficiency in terms of performance, delay, power, and area~\cite{xu2015approximate}, specifically for error-resilient applications. 
Recent investigations have shown that approximate computing can enable significant gains for error-tolerant applications, such as multimedia, image processing, deep learning, and data mining, which do not necessarily need full-precision output~\cite{9165786}.
  
Adders are essential arithmetic circuits, as they are one of the fundamental building blocks of other arithmetic operations, such as multiplication, division, and subtraction. Hence, the approximation of adders may significantly improve the performance and energy/power efficiency of any given application at the cost of some accuracy loss. 
Research efforts in the field of approximate adders have been directed toward designing efficient approximate adders that can offer better quality-efficiency trade-offs~\cite{jiang2015comparative,9165786}.
Note that the efficiency can be gauged based on essential evaluation metrics, including power, area, or latency (critical-path delay), depending on the user's preference. 
Generally, these metrics increase rapidly with the increase in the bit-width ($N$) of adders.

In general, state-of-the-art approximate adders are categorized into two main categories, i.e., low-latency approximate adders (LLAAs) and low-power approximate adders (LPAAs)~\cite{ayub2020peal}. LLAAs offer better delay characteristics as they trade accuracy for latency improvements by employing multiple sub-adder modules with smaller carry-chain lengths than the original design~\cite{ebrahimi2019block}. 
Almost Correct Adder (ACA)~\cite{verma2008variable}, Gracefully Degrading Adder (GDA)~\cite{ye2013reconfiguration}, Generic Accuracy Configurable Adder (GeAr)~\cite{shafique2015low}, Carry Cut-Back Adder (CCBA)~\cite{camus2018design} and Error Tolerant Adders (ETAs)~\cite{zhu2009design}\cite{zhu2010enhanced}\cite{zhu2011ultra} are a few examples of LLAAs.
The sub-adder modules in LLAAs can be disjoint or overlapping depending on the type and configuration of the LLAA. 
Each sub-adder contains some \textit{Resultant} bits ($R$ bits), which produce sum bits, and (optionally) some \textit{Prediction} bits ($P$ bits), which predict carry-in for the resultant part. ACA~\cite{verma2008variable}, ETA-I, ETA-II~\cite{zhu2009design}, ETA-IIM~\cite{zhu2011ultra} and ETA-III~\cite{zhu2010enhanced}  offer very restricted design space, as their $R$ and $P$ values are defined based on the type of the adder and the user-defined sub-adder length. 
To address this limitation, GDA~\cite{ye2013reconfiguration} and GeAr~\cite{shafique2015low} designs have been proposed. 
GDA employs disjoint modules of equal length, where each module is composed of an adder unit, responsible for computing the sum bits, and a carry-in prediction unit, responsible for predicting the carry-in for the subsequent module. 
Moreover, it employs multiplexers to offer run-time reconfigurability, where each multiplexer is responsible for selecting carry-in for a module either from its previous adder unit or from its previous carry-in prediction unit. 
Unlike GDA that offers run-time reconfigurability, GeAr is a configurable adder model that covers an extended design space of LLAAs, as it allows $R$ and $P$ to have any values given $R+P \leq N$. 
However, note that, even in GeAr, all sub-adders must have the same $R$ and $P$ values. 
To overcome this limitation, Quality-area optimal low-latency approximate Adder (QuAd)~\cite{hanif2017quad} proposed a model that allows each sub-adder to have any number of $R$ and $P$ bits regardless of the number of $R$ and $P$ bits in other sub-adders.
The analysis in QuAd showed that, given a latency constraint, it is possible to effortlessly select the optimal LLAA configuration from the whole design space of LLAAs. However, QuAd overlooks a predominant class of approximate adders, i.e., LPAAs, which may offer a better quality-efficiency trade-off.

Contrary to LLAAs, LPAAs are focused on offering better power/energy efficiency, which is mainly achieved through logic simplification of the underlying modules. IMPACT designs~\cite{gupta2011impact}, Low-power digital signal processing using approximate adders~\cite{6387646}, Inexact designs for approximate low power addition by cell replacement~\cite{7459392}, and XOR/XNOR-based approximate adders (AXA)~\cite{6720793} are a few of the well-known approximate adder designs that fall under the LPAA category. 

\textbf{Key Limitations and Associated Challenges:} The following points highlight the key limitations of state-of-the-art works and also present the associated challenges towards identifying/designing a superior class of approximate adders that can offer better quality-efficiency trade-off than conventional LLAAs as well as LPAAs.

\begin{itemize}
    \item $QuAd$~\cite{hanif2017quad} claims that adders composed of disjoint sub-adders of equal length, specifically $QuAd_o$ configurations, offer the best quality-latency trade-off out of all the LLAAs. Moreover, LPAAs are known to offer better quality for power/area efficiency trade-off. Although both LLAAs and LPAAs have been widely explored in the literature, hybrid designs that offer better latency as well as power and area characteristics without significant accuracy degradation, have not been explored. Towards this, it is important to identify the class and configurations of adders that can offer superior results to other predominant approximate adders. 
    \item Analyses in works like PEMACx~\cite{hanif2020pemacx} have highlighted that, based on the given scenario, a specific set of configurations can dominate the complete design space of LPAAs. Therefore, it is important to identify the LPAA configurations that can offer better results than all other LPAA designs under the given conditions and help construct optimal hybrid approximate adders. 
    \item Selecting the most efficient configuration, which offers the lowest area, power, and delay while meeting the user-defined accuracy constraints, is a challenging design space exploration problem, specifically when the number of potential configurations is huge. To select the most efficient configuration for a pre-defined accuracy constraint, different adder configurations have to be compared. However, efficient exploration requires fast yet accurate analytical models to estimate the quality as well as efficiency metrics of approximate adders. Therefore, such analytical models would be necessary for the newly identified class of hybrid approximate adders as well. 
\end{itemize}

\begin{figure}
\begin{minipage}[b]{0.62\textwidth}
\includegraphics[scale=0.2]{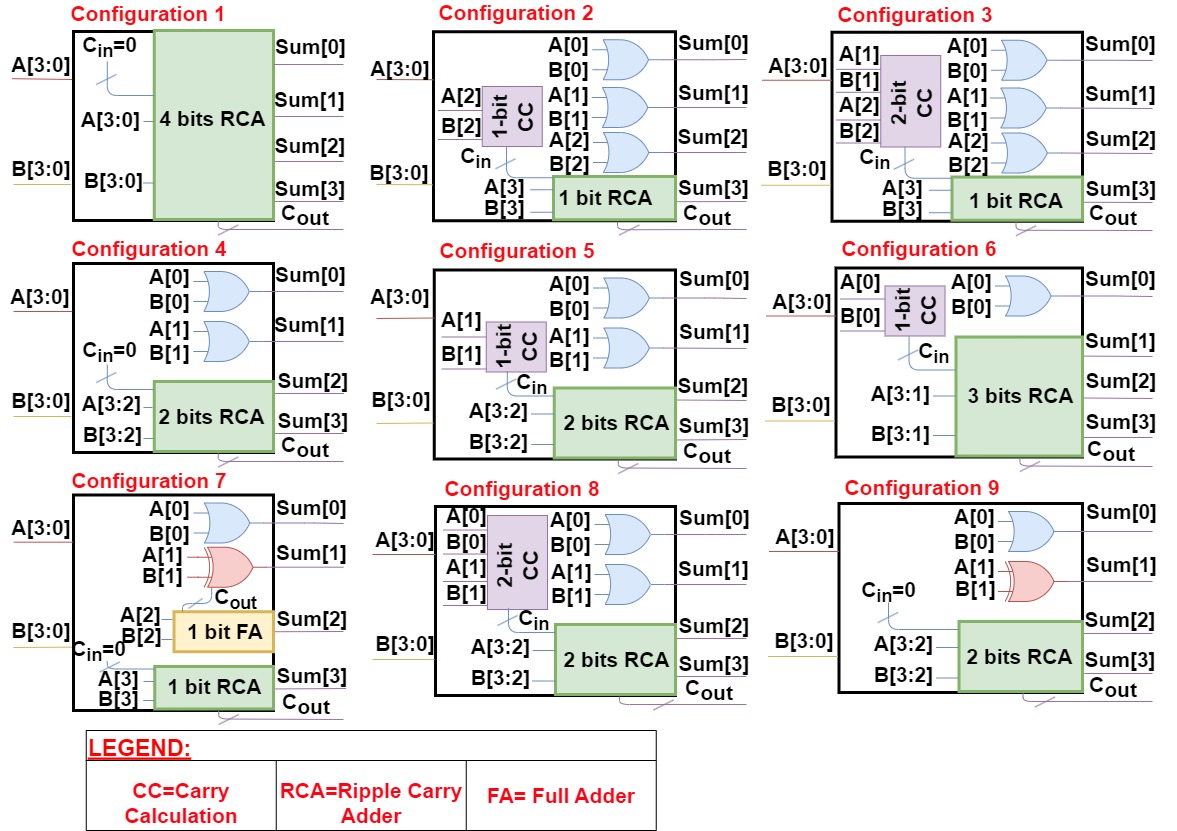}\\
\subcaption{}
\label{differentconf}
\end{minipage}%
\begin{minipage}[b]{0.33\textwidth}
\includegraphics[scale=0.5]{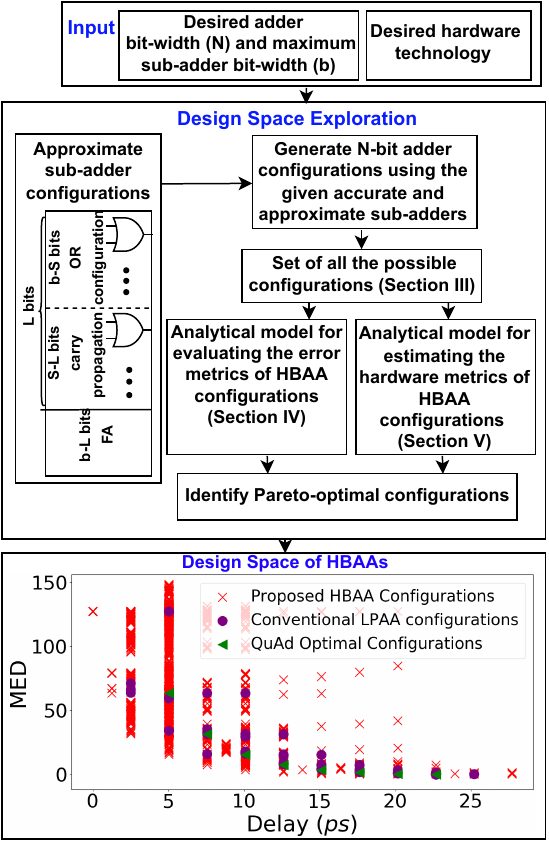}\\
\subcaption{}
\label{system diagram}
\end{minipage}%
\caption{Novel contributions. (a) A few of the proposed configurations for a 4-bit approximate adder block. (b) The flow of the proposed concepts for generating and selecting HBAA configurations.}
\end{figure}
\begin{figure}
    \centering
    \includegraphics[scale=0.5]{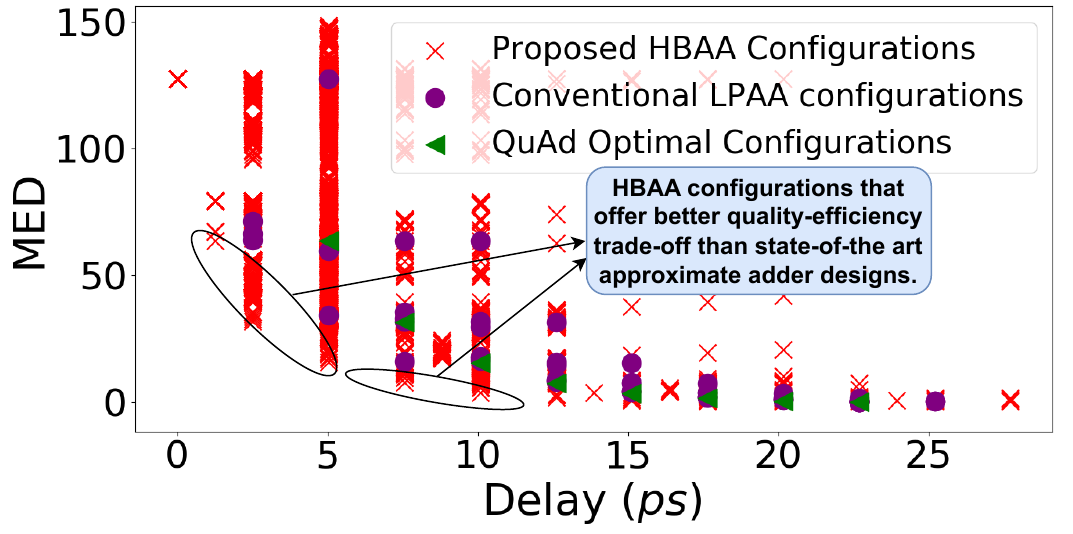}
    \caption{Design space of an 8-bit approximate adder}
    \label{design_space}
\end{figure}

\textbf{Overview of Our Novel Contributions}:
This paper focuses on building hybrid approximate adder designs that can offer better latency, power and area characteristics than conventional LLAAs and LPAAs. 
Considering the analysis in $QuAd$~\cite{hanif2017quad}, we focus on disjoint block-based approximate adders to achieve optimal quality-latency trade-off, while to achieve high power and area gains, we employ logic simplification concepts from LPAAs. 
As replacing the Full-Adders (FAs) at the least-significant locations with approximate variants have the least impact on the accuracy, 
we consider all the configurations in which the least significant FAs in each sub-adder block are replaced with approximate FAs as a part of our new design space. 
We assume that each sub-adder can have a different number of bits approximated, regardless of the number of bits approximated in other sub-adders. 
We mainly use OR gate-based approximations, i.e., replacing FAs with simple OR gates. 
Moreover, we allow arbitrary carry prediction length within sub-adders to predict the carry-in for accurate FAs present at the significant locations. 
As each sub-adder block in the proposed designs can have a different configuration (different from other sub-adders in the adder), we call these Heterogeneous Block-based Approximate Adders (HBAAs). Figure~\ref{differentconf} shows some of the possible configurations for 4-bit sub-adder blocks that can be used to construct larger HBAAs. 
Figure~\ref{system diagram} shows how such configurations can be combined to generate the complete design space of HBAAs. 
To show the superiority of the proposed configurations over the state-of-the-art adders, Figure~\ref{design_space} plots the complete set of 8-bit HBAAs over $QuAd_o$ configurations and LPAA configurations generated using the designs presented in~\cite{6387646} and~\cite{7459392}. The figure clearly shows that various HBAA configurations offer better results than $QuAd_o$ and conventional LPAA configurations. 
Note, for these results, we used Mean Error Distance (MED) as the main quality metric. 

\textbf{Key Novel Contributions}:
Figure~\ref{system diagram} presents our novel contributions in the form of a flow. The contributions are summarized as follows:

\begin{itemize}
\item We propose a new class of approximate adders called Heterogeneous Block-based Approximate Adders (HBAAs) that can offer better latency, power and area characteristics than conventional LLAA and LPAA designs. These adders mainly employ disjoint sub-adders to offer better quality-latency trade-off and logic simplifications in FAs to achieve higher area and power efficiency. For logic simplification, we replace FAs with OR gates, as they offer the best quality-efficiency trade-off when it comes to logic simplifications. 
\item We propose a generic accuracy-configurable adder model to represent HBAA configurations. The model enables us to build analytical models that can easily be used to estimate the error and hardware characteristics of HBAA configurations. 
\item We also present an analytical model for efficiently computing the PMF of error of HBAA configurations. The model facilitates convenient comparison of different adder configurations without requiring time-consuming and resource-hungry Monte-Carlo simulations, and thereby enables fast design space exploration of HBAA designs.
Apart from the analytical model for error estimation, we also present analytical models for estimating the delay, power and area characteristics of HBAA designs.
\end{itemize}

\textbf{Paper Organization}: The remainder of the paper is organized as follows: Section~\ref{Sec:Related_Work} provides a brief overview of approximate adders. Then, 
Section~\ref{sec:generic_model} presents a generic model for representing HBAAs. 
The proposed methodology for computing the PMF of error of HBAA configurations is presented in Section~\ref{Sec:Analytical_Model}. 
Section~\ref{Sec:hardware_metric_models} then presents the analytical models for estimating hardware metrics of HBAA configurations. 
Towards the end, Section~\ref{Sec:Results} presents the results of the proposed methodology and Section~\ref{Sec:Conclusion} concludes the paper.

\section{Related Works}
\label{Sec:Related_Work}
Approximate adder designs span a wide range of research efforts, i.e., from circuit level all the way to architectural level. In the earlier approaches, researchers mainly focused on transistor-level modifications to approximate adder circuits~\cite{gupta2011impact}\cite{gupta2012low}\cite{prabakaran2018demas}. 
Over time techniques such as voltage over-scaling (VOS)~\cite{karakonstantis2011voltage}-\cite{krause2011adaptive} and clock gating~\cite{kim2016designing} have also been employed to approximate circuits.  
However, the most prominent works in designing approximate adders are based on architectural-level modifications. 

As discussed in Section~\ref{sec:intro}, approximate adders can be classified into two main categories, i.e., LPAAs and LLAAs. 

\textbf{Low-Power Approximate Adders (LPAAs):} 
The primary approximate adder designs that fall in this category are IMPACT adders~\cite{gupta2011impact}\cite{6387646}, which are generated by simplifying the FA by reducing the number of transistors.   
Recently, researchers have focused on designing LPAAs through gate-level and architecture-level modifications. 
The approximate adders that fall in this category are inexact designs for approximate low-power addition by cell replacement~\cite{7459392} and approximate XOR/XNOR-based adders for inexact computing (AXA)~\cite{6720793}. 
Truth tables of some of the widely used LPAAs are presented in Table~\ref{tab:tabtruth}, where Types 1-5 correspond to IMPACT designs~\cite{gupta2011impact}\cite{6387646} achieved through transistors-reduction technique while Types 6-7 correspond to the inexact designs in~\cite{7459392} achieved through gates-reduction technique. 

\textbf{Low-latency Approximate Adders (LLAAs):}
A few adder designs that fall under the LLAAs category are: Almost Correct Adder (ACA-I)~\cite{verma2008variable}, Carry-Skip Approximate Adder (CSAA)~\cite{kim2013energy} and Gracefully Degrading Adder (GDA)~\cite{ye2013reconfiguration}. 

\begin{table}[ht]
    \centering
    \caption{Truth table of state-of-the-art LPAAs. The erroneous output are marked as red.}
    \adjustbox{width=\textwidth}{%
        \begin{tabular}{|c|c|c|c|c|c|c|c|c|c|c|c|c|c|c|c|c|c|c|c|c|c|c|c|c|c|c|}
        \hline
        \multicolumn{3}{|c|}{\textbf{Inputs}}&
        \multicolumn{3}{|c|}{\textbf{Accurate FA}}&
        \multicolumn{3}{|c|}{\textbf{LPAA Type 1}} &
        \multicolumn{3}{|c|}{\textbf{LPAA Type 2}}&
        \multicolumn{3}{|c|}{\textbf{LPAA Type 3}}&
        \multicolumn{3}{|c|}{\textbf{LPAA Type 4}}&
        \multicolumn{3}{|c|}{\textbf{LPAA Type 5}}&
        \multicolumn{3}{|c|}{\textbf{LPAA Type 6}}&
        \multicolumn{3}{|c|}{\textbf{LPAA Type 7}}  \\
        \hline
        $A$ & $B$ & $C_{in}$ & $Sum$ & $C_{out}$ & $Error$ & $Sum$ & $C_{out}$ & $Error$& $Sum$ & $C_{out}$ & $Error$ &$Sum$ & $C_{out}$ & $Error$&$Sum$ & $C_{out}$ & $Error$&$Sum$ & $C_{out}$ & $Error$&$Sum$ & $C_{out}$ & $Error$&$Sum$ & $C_{out}$ & $Error$\\ 
        \hline
        $0$ & $0$ & $0$ 
        & $0$ & $0$ & $0$ 
        & $0$ & $0$ & $0$
        & $\textcolor{red}{1}$ & $\textcolor{red}{0}$ & $\textcolor{red}{1}$
       & $\textcolor{red}{1}$ & $\textcolor{red}{0}$ & $\textcolor{red}{1}$
        & $0$ & $0$ & $0$
        & $0$ & $0$ & $0$
        & $0$ & $0$ & $0$
        & $0$ & $0$ & $0$\\ 
        \hline
        $0$ & $0$ & $1$ 
        & $1$ & $0$ & $0$ 
        & $1$ & $0$ & $0$
        & $1$ & $0$ & $0$
        & $1$ & $0$ & $0$
        & $1$ & $0$ & $0$
        & $\textcolor{red}{0}$ & $\textcolor{red}{0}$ & $\textcolor{red}{-1}$
        & $\textcolor{red}{1}$ & $\textcolor{red}{1}$ & $\textcolor{red}{2}$
        & $1$ & $0$ & $0$\\ 
        \hline
        $0$ & $1$ & $0$ 
        & $1$ & $0$ & $0$ 
        & $\textcolor{red}{0}$ & $\textcolor{red}{1}$ & $\textcolor{red}{1}$
        & $1$ & $0$ & $0$
        & $\textcolor{red}{0}$ & $\textcolor{red}{1}$ & $\textcolor{red}{1}$
        & $\textcolor{red}{0}$ & $\textcolor{red}{0}$ & $\textcolor{red}{-1}$
        & $1$ & $0$ & $0$
        & $1$ & $0$ & $0$
        & $1$ & $0$ & $0$\\  
        \hline
        $0$ & $1$ & $1$ 
        & $0$ & $1$ & $0$ 
        & $0$ & $1$ & $0$
        & $0$ & $1$ & $0$
        & $0$ & $1$ & $0$
        & $\textcolor{red}{1}$ & $\textcolor{red}{0}$ & $\textcolor{red}{-1}$
        & $\textcolor{red}{1}$ & $\textcolor{red}{0}$ & $\textcolor{red}{-1}$
        & $0$ & $1$ & $0$
        & $\textcolor{red}{1}$ & $\textcolor{red}{1}$ & $\textcolor{red}{1}$\\  
        \hline
        $1$ & $0$ & $0$ 
        & $1$ & $0$ & $0$ 
        & $\textcolor{red}{0}$ & $\textcolor{red}{0}$ & $\textcolor{red}{-1}$
        & $1$ & $0$ & $0$
        & $1$ & $0$ & $0$
        & $\textcolor{red}{0}$ & $\textcolor{red}{1}$ & $\textcolor{red}{1}$
        & $\textcolor{red}{0}$ & $\textcolor{red}{1}$ & $\textcolor{red}{1}$
        & $1$ & $0$ & $0$
        & $1$ & $0$ & $0$\\ 
        \hline
        $1$ & $0$ & $1$ 
        & $0$ & $1$ & $0$ 
        & $0$ & $1$ & $0$ 
        & $0$ & $1$ & $0$ 
        & $0$ & $1$ & $0$ 
        & $0$ & $1$ & $0$ 
        & $0$ & $1$ & $0$ 
        & $0$ & $1$ & $0$ 
        & $\textcolor{red}{1}$ & $\textcolor{red}{1}$ & $\textcolor{red}{1}$\\ 
        \hline
        $1$ & $1$ & $0$ 
        & $0$ & $1$ & $0$ 
        & $0$ & $1$ & $0$ 
        & $0$ & $1$ & $0$ 
        & $0$ & $1$ & $0$ 
        & $0$ & $1$ & $0$
        & $\textcolor{red}{1}$ & $\textcolor{red}{1}$ & $\textcolor{red}{1}$
        & $\textcolor{red}{0}$ & $\textcolor{red}{0}$ & $\textcolor{red}{-2}$
        & $0$ & $1$ & $0$\\ 
        \hline
        $1$ & $1$ & $1$ 
        & $1$ & $1$ & $0$ 
        & $1$ & $1$ & $0$
        & $\textcolor{red}{0}$ & $\textcolor{red}{1}$ & $\textcolor{red}{-1}$
        & $\textcolor{red}{0}$ & $\textcolor{red}{1}$ & $\textcolor{red}{-1}$
        & $1$ & $1$ & $0$
        & $1$ & $1$ & $0$
        & $1$ & $1$ & $0$
        & $1$ & $1$ & $0$\\ 
        \hline
       \end{tabular}}%
    \label{tab:tabtruth}%
\end{table}%

\begin{figure}
    \centering
    \subfloat[LPPA type 5]{
        \includegraphics[width=0.35\columnwidth]{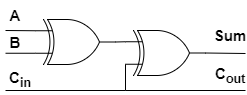}
        \label{gate LPPA 1}
    }\hfil
    \subfloat[LPPA type 6]{
         \includegraphics[width=0.35\columnwidth]{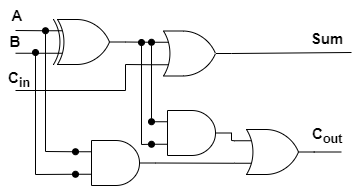}
        
         \label{gate LPPA 2}
    }
      \caption{An example of gate reduction LPPAs}
      \label{gate LPPA}
\end{figure}{}

Apart from the above-mentioned adder designs, Zhu \textit{et al.} proposed four different variants of error-tolerant LLAAs, i.e, ETA~I and ETA~II~\cite{zhu2009design}, ETA~III~\cite{zhu2010enhanced}, ETA~IIM~\cite{zhu2011ultra}.  
Another LLAA has been proposed in~\cite{ebrahimi2019block} for energy-efficient applications. In this design, the non-overlapping sub-adders use a carry predictor unit and a selector unit to decide whether the carry-out of each sub-adder is propagated or not. 
Generic Accuracy Configurable Adder (GeAr) is proposed in~\cite{shafique2015low}, which utilizes redundant blocks leading to excessive hardware overhead. 
The Reverse Carry Propagate Adder (RCPA)~\cite{pashaeifar2018approximate} propagates the carry-in in a counter-flow manner from the MSB to the LSB. The RCPA is not efficient in terms of energy. 
In general, it should be noted that these methods have fixed configurations with limited flexibility and a massive error value.

On the other hand, some methods contribute to flexibility in design by supporting multiple configurations. QuAd~\cite{hanif2017quad} is an enhanced model of GeAr with flexibility. 
In the QuAd adder, the sub-adders can have different sizes as well as different carry prediction lengths. A reconfigurable approximate adder is proposed in~\cite{akbari2016rap}, and it employs the carry look ahead (CLA) method. The adder is split into two disjoint segments, i.e., the approximate part and the augment part. The adder design enables the user to switch between accurate and approximate operations by using a multiplexer. This technique imposes some hardware overheads. Xu~\textit{et al.} proposed another reconfigurable adder called Simple Accuracy-Reconfigurable Adder (SARA) \cite{xu2018simple}. In SARA, the adder is divided into $K$ disjoint sub-adders. Moreover, it uses an error recovery circuit to reduce the error, which imposes additional hardware overheads. Additionally, the utilized ripple carry adder in the sub-adder causes a long critical path. Note that in \cite{hanif2017quad}, \cite{akbari2016rap} and \cite{xu2018simple}, the flexibility offered by design comes with additional hardware cost, and these designs only consider homogeneous blocks. 

\textbf{Analytical Models for Error Estimation:} For selecting the most efficient design for a given application, we need to conduct a comparative analysis that takes into account error metrics, critical path delay, design area, and energy consumption. Error metrics analysis is typically performed using computer simulations. However, as the size of the adder increases, the exhaustive simulation time increases exponentially. So, the exhaustive simulation technique becomes time-consuming and thus impractical. Therefore, research efforts have been directed towards proposing analytical models to facilely assess error metrics of different types of approximate adders. An analytical model for homogeneous overlapping blocks is proposed in~\cite{mazahir2016probabilistic}. In addition to proposing a generic methodology for error probability estimation, the paper also presented a method to evaluate the PMF of error value. 
Another analytical model is proposed in~\cite{celia2018probabilistic}. The paper focuses on error metrics of adders with two segments, one accurate and the other inaccurate segment. 
In~\cite{dutt2018analysis}, error metrics are obtained based on an analytical model and generalized analytical model for equal redundant segments with homogeneous blocks. 
Moreover, the authors have used an optimization technique to optimize the design's estimated parameters such as delay, power, and area. 
In the optimization framework, the given accuracy is considered a hard constraint. 
However, the drawback of these analytical methods is that they do not consider heterogeneous approximate blocks in the precise evaluation of the error probability of approximate adders. 

Moreover, an analytical model for error metrics, e.g., ER and MSE, of low-power approximate adder is proposed by the PEAL~\cite{ayub2020peal}. 
It obtains the error metrics by evaluating the carry-out probability for each approximate FA. 
This method only evaluates the error rate as the accuracy of low-power approximate adder and cannot be used to estimate more relevant error metrics such as MSE, MED or PMF of error value. 
PEMACx~\cite{hanif2020pemacx} is a novel analytical method for efficiently computing the PMF of error of a low-power approximate adder that is composed of cascaded approximate adder units. 
In this article, the probability of carry-out error is evaluated for each cascaded approximate adder unit. 
These probabilities are used recursively as carry-in probabilities for the next stages to recursively evaluate the probability of carry-out error until the last stage. 
Also, \cite{8994186} proposed a fast analytical method to calculate the PMF of error value for low-latency and low-power approximate adders. 
These models are generalized to support multiple different types of low-latency and low-power approximate adder configurations.
Therefore, the computational time for calculating the MED can still be improved. 
For this purpose, in this work, we develop a specialized and more efficient analytical model to compute error metrics of approximate adder configurations that fall in the HBAA category. 
As the proposed model is specialized for HBAA adders, it takes less time to generate accurate estimates for HBAA configurations. 
In this regard, in the following sections, we first provide a generic model of our proposed HBAA configurations, then we provide an analytical model that is used to evaluate the PMF of error of HBAAs using statistical formulas derived from basic probability theory. 

\section{Generic Model for HBAA adders}
\label{sec:generic_model}

\begin{figure}
    \centering    \includegraphics[width=0.8\columnwidth]{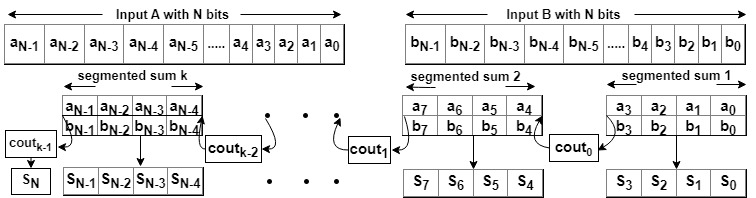}
     \caption{General diagram of an  adder composed of $k$ disjoint sub-adder blocks}
    \label{fig1}
\end{figure}

\begin{figure}
    \centering
    \includegraphics[width=0.6\columnwidth]{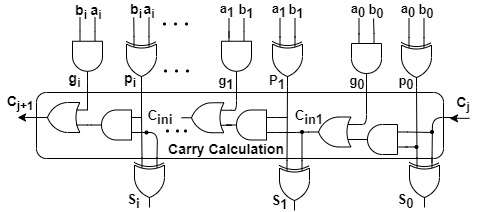}
    \caption{Logic expression of carry-out generation for RCA}
    \label{fig2}
\end{figure}

An HBAA adder operates on two N-bit inputs  $A=(a_{N-1},a_{N-2},...,a_i,...,a_0)$ and $B=(b_{N-1},b_{N-2}\\,...,b_i,...,b_0)$. It is mainly composed of $k$ disjoint sub-adder blocks, as illustrated in Figure~\ref{fig1}.
{}
First, we explain the conventional Ripple Carry Adder (RCA) and then our modifications that lead to a new design space. In an accurate adder, the carry-out at $i^{th}$ bit location is calculated using Eq.~\ref{eq1}. The equation is based on generate and propagate signals of the previous bit locations. The generate and propagate signals are computed using Eq.~\ref{eq2} for each $i^{th}$ bit location, where $i \in \{0, 1, 2, ..., N-2, N-1\}$.
\begin{equation}
c_{i+1}=g_i+p_i g_{i-1}+...+g_1 \prod_{j=2}^{i}p_j+g_0 \prod_{j=1}^{i}p_j+c_i\prod_{j=0}^{i}p_j
\label{eq1}  
\end{equation} 
\begin{equation}
p_i=a_i\oplus b_i, g_i=a_i.b_i
\label{eq2}
\end{equation}  
Here, $c_i$ represents carry-in and $c_{i+1}$ represents carry-out of $i^{th}$ bit location. $g_i$ and $p_i$ correspond to generate and propagate signals of the $i^{th}$ bit location, respectively. The carry-out logic circuit is illustrated in Figure~\ref{fig2}. 
{}

\begin{figure}
    \centering
    \includegraphics[width=0.3\columnwidth]{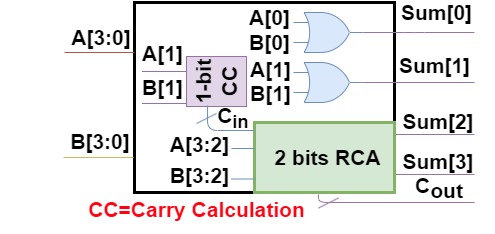}
    \caption{An example of approximate block with $H=4$ bits length}
    \label{fig3}
\end{figure}

The proposed adder is composed of $k$ blocks where each block is an $H$-bit sub-adder and $k=\lceil N/H \rceil $. 
The blocks used in this adder are not homogeneous and fall into two types, i.e., accurate and approximate blocks. 
The accurate blocks are used primarily at MSB locations and the approximate blocks are used at LSB locations. 
The accurate blocks are based on the Ripple Carry Adder (RCA), and the approximate blocks use a combination of logic simplification (OR gates replacement), full adders, and the RCA design to perform the addition of the corresponding bits.
For computing the carry-out signal of an approximate block, any carry-chain length can be selected. 
Consequently, we are free to define the length of carry generation and propagation in every approximate block. 
For instance, Figure~\ref{fig3} shows a 4-bit approximate block with carry propagation length equals three and the lower two FAs replaced with OR gates for computing the corresponding sum bits.
{}

\begin{figure}
    \centering
    \includegraphics[width=0.3\columnwidth]{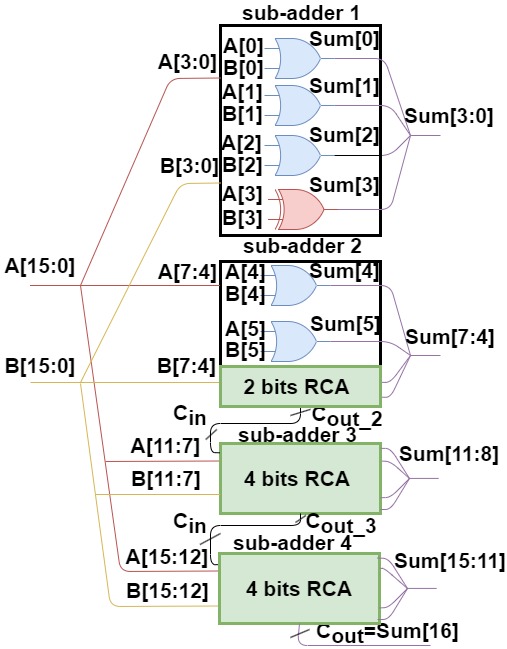}
    \caption{A 16 bits HBAA configuration composed of 4-bit sub-adder blocks}
    \label{fig4}
\end{figure}

We can have heterogeneous approximate blocks with different configurations placed in different positions in the proposed adder. 
Moreover, we also consider that carry propagation occurs between the most significant approximate block and all the accurate blocks on the most significant side in the adder. This is illustrated in Figure~\ref{fig4} as well for a 16-bit HBAA configuration composed of 4-bit sub-adder blocks. As can be seen in the figure, the carry-out of the most significant approximate block (i.e., sub-adder 2) is connected to the carry-in of the next block (i.e., sub-adder 3) and all the accurate blocks on the most significant side (i.e., sub-adder 3 and sub-adder 4) are also connected.

{}

For our HBAA, each approximate sub-adder can have any number of bits of inexact logic (OR gates) $L$ and any carry chain length $S$. An $N-bit$ HBAA adder consists of $k$ approximate sub-adders of equal size. The adder is defined using an inexact logic configuration vector, $L_{vec}=[L_1,L_2,...,L_k]$, and a carry chain vector, $S_{vec}=[S_1,S_2,...,S_k]$. Here, $L_i$ and $S_i$ represent the number of inexact logic bits and the carry-chain length of the $i^{th}$ sub-adder, respectively. Hence, the generic HBAA representation, $HBAA\{[L_1,L_2,...,L_k],[S_1,S_2,...,S_k]\}$, fully defines any possible HBAA configuration.

In the following section, we discuss the analytical model for computing PMF of error of HBAA configurations.

\section{Analytical Modeling for Computing Error Metrics}
\label{Sec:Analytical_Model}

Besides the conventional performance metrics such as delay, area, and power, error metrics are also important to compare different approximate adder configurations and designs.  
Metrics such as Error Distance (ED), Mean Error Distance (MED)~\cite{6569370}\cite{shafique2015low}\cite{6226361}\cite{6797866}\cite{mazahir2016probabilistic}, Normalized Mean Error Distance (NMED)~\cite{shafique2015low}\cite{6226361}, and Error Rate (ER)~\cite{ebrahimi2019block} are commonly used to quantify the computational accuracy/quality of approximate arithmetic circuits. 
Among these metrics, the error distance (ED), and mean error distance (MED) are considered more important and applicable for the comparison of approximate adders \cite{wu2018efficient}. 
These metrics can be calculated either using computer simulations or analytical models. 
However, due to the greater benefits of analytical models over computer simulations in terms of execution time/cost, analytical models are preferred for quality and performance estimation of approximate components, specifically for design space exploration tasks. 
Therefore, in this section, we present a novel analytical model for estimating error metrics of HBAA configurations. 
The primary advantages of the proposed analytical model consist of:
\begin{itemize} 
  \item It facilitates efficient comparison between different HBAA configurations. 
  \item It can be used to explore the complete design space of HBAA configurations in order to obtain optimal circuit parameters such as the carry chain length, the number of approximate blocks, and the configuration of each approximate block. 
\end{itemize}

In the following text, we introduce our proposed analytical model for computing the PMF of error of an HBAA configuration. 
The PMF of error indicates all possible error values and the probability of each error value.
It is important as it can be used to compute most of the error metrics such as maximum absolute error value, MED, NMED, MSE, and error probability. 
Moreover, it also presents an estimate of the distribution of the error, not just the mean values. 
As an HBAA adder is composed of multiple sub-adder blocks, the error in each approximate block can propagate to the adder's output.  
The sources of error in the adder's output are errors in the carry chain due to truncation and approximation errors in the internal computations of each block due to the replacement of FAs with OR gates for sum generation. 
To evaluate the PMF of error value of an HBAA configuration, first, we identify the sources of errors in approximate blocks. 
Next, we evaluate the PMF of error of each approximate block independently. 
Eventually, we combine the PMFs of the blocks to get the overall PMF of error of the HBAA configuration. 
The proposed methodology is shown in figure~\ref{figmethod}, which consists of the following stages:
\begin{figure}
    \centering
    \includegraphics[scale=0.6]{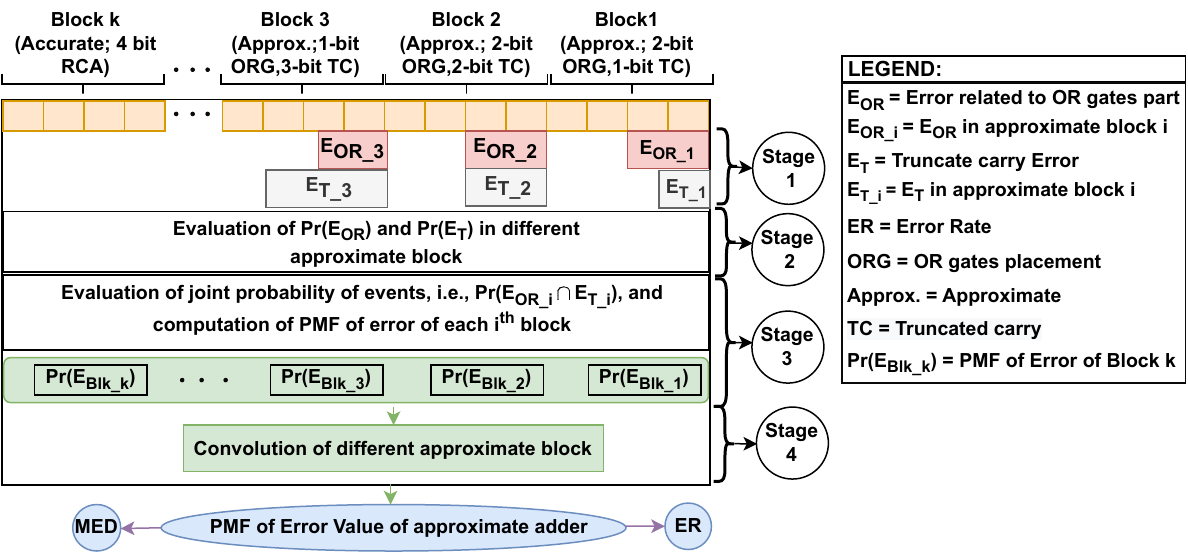}
    \caption{Proposed methodology for computing the PMF of error value of HBAA}
    \label{figmethod}
\end{figure}{}

\begin{itemize}
    \item \textbf{Identification of error sources (Stage~1):} The first stage is for identifying the error sources in the approximate blocks of the given HBAA configuration. The errors related to replacement of FAs with OR gates for sum computation are referred to as $E_{OR}$, and the errors related to carry-chain truncation are referred to as $E_{T}$.  
    \item \textbf{Evaluation of the PMF of each error source (Stage~2):} The second stage is for computing the PMF of $E_{OR}$ and $E_T$ error types in each sub-adder block independently. The analytical models of these errors are presented in Section~\ref{section_origin_of_err}. 
    \item \textbf{Evaluation of the PMF of error of each approximate block (Stage 3):} In this stage, an analytical model is proposed to find the joint error events in each approximate block. Then, the PMF of error value of each approximate block is obtained by using the probability of corresponding error sources and the corresponding joint probability of events. The details of this stage are presented in Section~\ref{sec:evaluationapp}.
    \item \textbf{Evaluation of PMF of error of the complete HBAA configuration (Stage 4):} The PMF of error value of the complete HBAA configuration is calculated by using independent error events of all the sub-adder blocks. Thus, in this case, it is computed by convolving the PMF of error value of all the approximate blocks. The details of this stage are presented in Section~\ref{sec_approx_block_total}.
\end{itemize}

\subsection{Identification of the Error Sources and Evaluation of Their PMFs}
\label{section_origin_of_err}

In HBAAs, we consider two different types of approximations that can lead to errors in the adder's output, and we identify them as two separate error sources. 
The first type is replacement of FAs with OR gates and the second is carry-chain truncation. 
In this work, we refer the errors related to replacement of FAs with OR gates as $E_{OR}$, and the errors related to carry-chain truncation as $E_T$. 
The computation of PMFs of $E_{OR}$ and $E_T$ for each approximate sub-adder block (i.e., Stage~2 in Figure~\ref{figmethod}) is explained in the following sub-sections. 

\subsubsection{Evaluation of PMF of $E_{OR}$ for Each Approximate Block} 
When $L$ least significant FAs of an adder block are replaced with OR gates, the error value can range from $0$ to $2^L-1$. 
Assuming all the input bits to be independent, the probability of each possible error value can be computed by using the probabilities of error at individual bit locations where the FAs are replaced with OR gates, as an OR gate leads to either $0$ or $1$ error at the corresponding bit location. 
The error value at a given bit location $i$ is $1$ when both the input bits at the corresponding bit location are $1$, and error value is $0$ when at least one of the input bits is $0$. 
Hence, assuming $Pr(a_i = 1)$ corresponds to the probability of the $i^{th}$ bit of input $A$ being $1$ and $Pr(b_i = 1)$ corresponds to the probability of the $i^{th}$ bit of input $B$ being $1$, the probability of the error value being $1$ can be computed using the probability of the generate signal of the corresponding location. 
Given, $g_i$ represents the generate signal at $i^{th}$ bit location, the probability of error value at $i^{th}$ bit location being $1$ (when the FA at the corresponding location is replaced with an OR gate) can be computed using the following equation. 

\begin{equation}
    Pr(g_i=1)=Pr(a_i=1)\cap Pr(b_i=1)
    \label{eq:genrate}
\end{equation} 

Since all the input bits are assumed to be independent of each other, the intersection can be replaced with the product of the two probabilities. Thus, Eq.~\ref{eq:genrate} can be simplified to: 

\begin{equation}
    Pr(g_i=1)=Pr(a_i=1). Pr(b_i=1)
    \label{eq:generate_simple}
\end{equation}

Similarly, the probability of error value being $0$ at the same bit location can be computed using $1 - Pr(g_i=1)$.

\begin{figure}
    \centering    \includegraphics[scale=0.6]{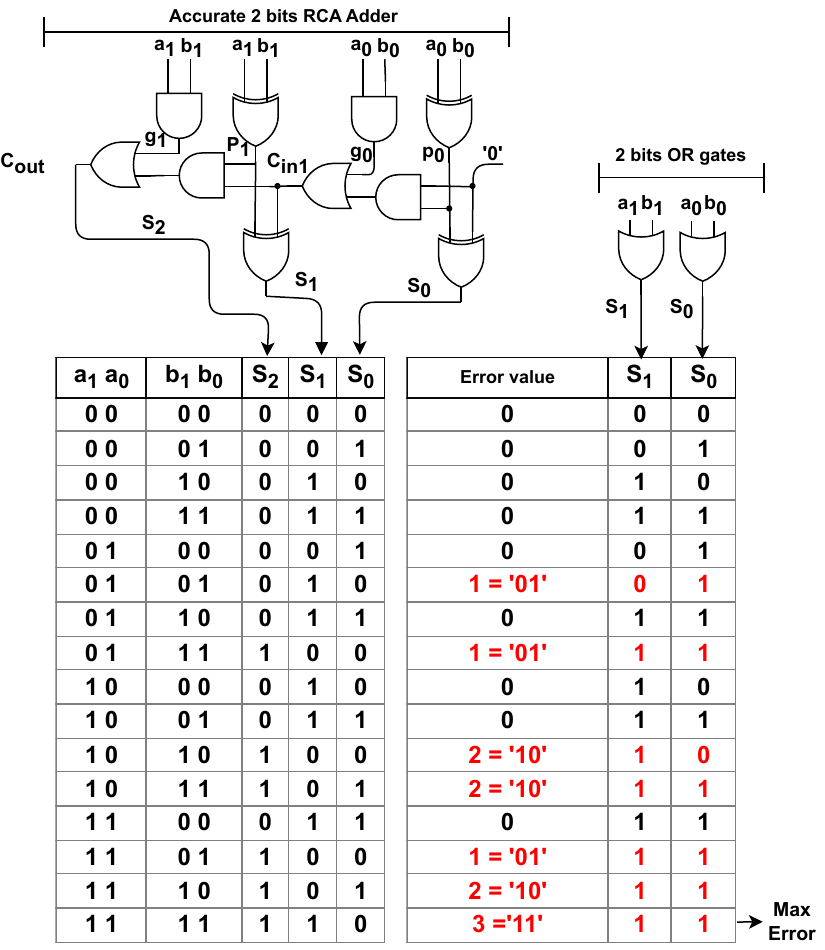}
    \caption{A comparison of the truth tables of an accurate 2-bit adder composed of two FAs with an approximate adder composed of two OR gates. The error cases are marked in red.}
    \label{fig5}
\end{figure}{}

\textbf{Example for Computing PMF of $E_{OR}$ for a 2-bit Adder:} Here, we present an example to demonstrate the usability of the above method for computing the PMF of $E_{OR}$ of a 2-bit approximate adder composed of two OR gates, shown on the right side of Figure~\ref{fig5}. 
For the two bit approximate adder, the error value range from $0$ to $3$. 
Figure~\ref{fig5} highlights all the error cases of the two bit approximate adder. 
For the considered case, the probability of each error value can be computed by using the binary representation of the error value. 
For example, for error value ($x$) equals $3$, by converting $3_{10}$ to its binary representation (i.e., $11_2$), we can compute its probability using the generate signals of the corresponding locations as shown in Eq.~\ref{EV_3_Case}. 

\begin{equation}
    Pr(x=3) = Pr(g_1 = 1) \cap Pr(g_0 = 1)
\label{EV_3_Case}
\end{equation}

Assuming the input bits to be independent of each other, the above equation can be written as: 

\begin{equation}
    Pr(x=3) = Pr(g_1 = 1).Pr(g_0 = 1)
\label{EV_3_Case_1}
\end{equation}

Following the same procedure, the PMF of $E_{OR}$ can be written as: 

\begin{equation}
    Pr(x)=
\begin{cases}
Pr(g_1 = 0).Pr(g_0 = 0) & \text{$x=0$} \\
Pr(g_1 = 0).Pr(g_0 = 1) & \text{$x=1$} \\
Pr(g_1 = 1).Pr(g_0 = 0) & \text{$x=2$} \\
Pr(g_1 = 1).Pr(g_0 = 1) & \text{$x=3$} \\
\end{cases}
\label{EV_PMF_Case}
\end{equation}

Assuming uniformly distributed inputs, $Pr(g_0 = 1)$ can be computed as: 

\begin{equation}
    Pr(g_0 = 1) = Pr(a_0 = 1).Pr(b_0 = 1) = \frac{1}{2}.\frac{1}{2} = \frac{1}{4}
\label{g_0_val}
\end{equation}

Similarly, we get: 

\begin{equation}
    Pr(g_1 = 1) = \frac{1}{4}
\label{g_1_val}
\end{equation}

By putting the values of $Pr(g_0 = 1)$ and $Pr(g_1 = 1)$ in Eq~\ref{EV_PMF_Case} while considering $Pr(g_0 = 0) = 1 - Pr(g_0 = 1)$ and $Pr(g_1 = 0) = 1 - Pr(g_1 = 1)$, we get: 

\begin{equation}
    Pr(x)=
\begin{cases}
\frac{9}{16} & \text{$x=0$} \\
\frac{3}{16} & \text{$x=1$} \\
\frac{3}{16} & \text{$x=2$} \\
\frac{1}{16} & \text{$x=3$} \\
\end{cases}
\label{EV_PMF_Case_Final}
\end{equation}

\textbf{Generalized Model for PMF of $E_{OR}$ for an $L$-bit Adder:} According to the above description, we can formulate the probability of each error value ($x$) of an $L$-bit approximate adder composed of $L$ OR gates using the following equation. 

\begin{equation}
     Pr(E_{OR}=x) = \prod_{i \in I} Pr(g_i=1) . \prod_{j \in J} Pr(g_j=0)
\label{eqsum_error}
\end{equation}

Here, $I$ represents the set of bit locations where generate signal is 1 and $J$ represents the set of bit locations where generate signal is 0. Now, assuming uniform distribution for the inputs, Eq.~\ref{eqsum_error} can be re-written as follows: 

\begin{equation}
     Pr(E_{OR}=x)=(\frac{1}{4})^{len(I)}
      . (\frac{3}{4})^{L-len(I)}
\label{eq:or_error}
\end{equation}

where, $len(I)$ represents the number of elements in the set $I$. 

\subsubsection{Evaluation of PMF of $E_T$ for Each Approximate Block}
\label{sec_carry_trunc}

To evaluate the PMF of $E_T$, we need a model for computing the distribution of the sum of two bit-level subsets of inputs to the adder. 
To define that, first, we define a model for computing the distribution of a subset of bits of an input based on \cite{mazahir2016probabilistic}. 
If $A_{sub}=[a_{q_2},...,a_{q_1}]$ is a sub-group of $n$ bits from $A=[a_{N-1},a_{N-2},...,a_0]$, where $0<q_1<q_2<N$ and $n=q_2-q_1+1$, we can derive the probability distribution of $A_{sub}$ (i.e., $P_{A_{sub}}(r)$ for $0 \leq r \leq 2^{q_2-q_1+1}-1$) as follows:

 \begin{equation}
    \begin{split}
      P_{A_{sub}}(r)=&\sum_{i=0}^{2^{N-1-q_2}-1}\bigg(\sum_{j=0}^{2^{q_1}-1}P_A(2^{q_2+1}i+2^{q_1}r+j)\bigg)\\
        &0 \leq r \leq 2^{q_2-q_1+1}-1
    \end{split}
 \label{eq8} 
 \end{equation}

Similarly, $P_{B_{sub}}$ can be derived from $P_B$ for the other input $B$. 
Since the two inputs are independent, the PMF of the summation $Z=A_{sub}+B_{sub}$ is calculated by convolving $P_{A_{sub}}$ with $P_{B_{sub}}$. 
Assuming that the probability distribution of $A$ and $B$ are uniform between $0$ and $2^N-1$, $A_{sub}$ and $B_{sub}$ can be considered uniform between $0$ and $2^n-1$. 
Therefore, the PMF of $Z$ can be represented as follows: 

\begin{subequations}
\begin{align}
&P_Z(r;n)=P_{A_{sub}}(r;n)\circledast P_{B_{sub}}(r;n)\\
&P_Z(r;n)=
\begin{cases}
\frac{r+1}{2^{2n}} & \text{$0\leq r \leq 2^n-1$} \\
\frac{2^{n+1}-r-1}{2^{2n}} & \text{$2^n-1\leq r \leq 2^{n+1}-2$} \\
0 & \text{otherwise}
\end{cases}
\end{align}
\label{eq9}
\end{subequations}

Carry chain can be truncated at any bit location inside an $H$-bit approximate block, as explained in Section~\ref{sec:generic_model}. 
If the length of the carry chain is $S$ bits, the length of the truncated portion is $H-S$ bits (shown in Figure~\ref{truncate_carry}), which can lead to an error of $2^{H-S}$ at the output of the block. 
The error occurs only when the first $H-S$ bit segment shown in Figure~\ref{truncate_carry} is in generate mode. 
Hence, the probability of $E_T = 2^{H-S}$ can be represented as: 

\begin{figure}[ht]
    \centering
    \includegraphics[width=0.4\columnwidth]{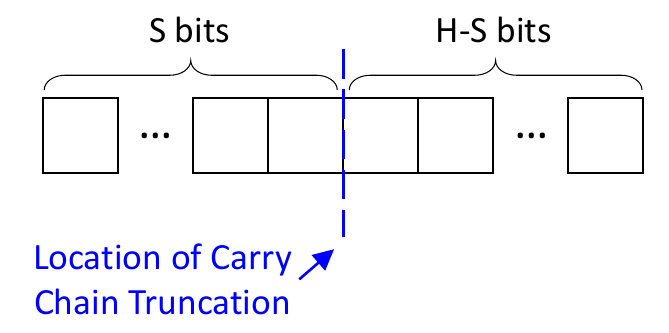}
    \caption{Approximate block with carry chain truncated at bit location H-S.}
    \label{truncate_carry}
\end{figure}{}

\begin{equation}
    Pr(E_T=2^{H-S})=Pr(G_1)
\label{truncate_carry_formul1}
\end{equation}

Where $G_1$ represents carry generation events of the first segment. Thus, the PMF of the error value of $E_T$ can be computed using the following equation.
\begin{equation}
    Pr(E_T=y)=
    \begin{cases}
     Pr(G_1) & \text{$y=2^{H-S}$} \\
     1- Pr(G_1) & \text{$y=0$} \\
    \end{cases}
\label{error_value_truncate}
\end{equation}

Note that an event in $G_1$ occurs when the summation of the corresponding input bits is at least $2^{H-S}$. 
Therefore, the probability of $G_1$ can be formulated as follows:
\begin{equation}
    Pr(G_1) = Pr(A_{sub}+B_{sub} > 2^{H-S}-1)
    =Pr(Z > 2^{H-S}-1)
 \label{generation_prob}
\end{equation}
which can be further expanded to Eq.~\ref{generation_prob_2}. 
\begin{equation}
    Pr(G_1)= \sum_{j=2^{H-S}}^{2^{H-S+1}-2}P_Z(j;H-S)
\label{generation_prob_2}
\end{equation} 
By substituting Eq.~\ref{generation_prob_2} in Eq.~\ref{error_value_truncate}, the PMF of $E_T$ can be given as:
\begin{equation}
    Pr(E_T=y)=
    \begin{cases}
     \sum_{j=2^{H-S}}^{2^{H-S+1}-2}P_Z(j;H-S) & \text{$y=2^{H-S}$} \\
     1 - \sum_{j=2^{H-S}}^{2^{H-S+1}-2}P_Z(j;H-S) & \text{$y=0$} \\
     0 & otherwise
    \end{cases}
\label{eqation1501}
\end{equation}

\subsection{Evaluation of PMF of Error of Individual Approximate Blocks} \label{sec:evaluationapp}

The method for computing the PMF of error of an approximate block of an HBAA configuration depends on the configuration of the block. 
Mainly, we divide the block configuration into three types based on the conditions listed in Table~\ref{tab:tab10}. 
A method for computing the PMF of error for each individual case is presented in the following text. 

\begin{table}[ht]
    \centering
    \caption{Error value ranges based on different cases}
    \adjustbox{width=0.8\columnwidth}{%
        \begin{tabular}{|c|c|c|}
        \hline
        \textbf{Case}  & \textbf{Carry chain truncated and OR gates position} & \textbf{Error value ranges} \\ 
        \hline
        \multirow{2}{*}{\textbf{First}} & \multirow{2}{*}{\normalsize$H-S > L$} & $0 \leq error \leq 2^{L}-1$\\ \cline{3-3} 
         &  & $2^{H-S} \leq error \leq (2^{H-S}) + (2^{L}-1)$\\ 
         \hline
        \textbf{Second} & \normalsize$H-S = L$ & $0 \leq error \leq 2^{L}-1$\\ 
        \hline
        \multirow{3}{*}{\textbf{Third}} & \multirow{3}{*}{\normalsize$H-S < L$} &  $0 \leq error \leq 2^{L}-1$\\ \cline{3-3} 
         &    & \multirow{2}{*}{ $error = X-2^{L} ,  2^{H-S}\leq X \leq 2^{L}-1 $ }\\   &  & \\ 
         \hline
       \end{tabular}}%
    \label{tab:tab10}%
\end{table}%

\begin{figure}
    \centering
    \includegraphics[scale=0.55]{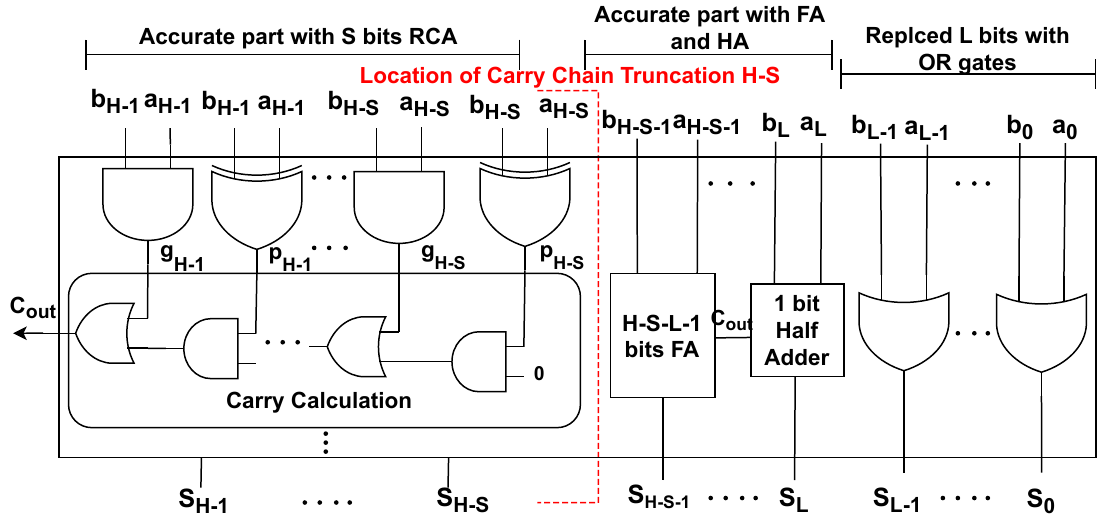}
    \caption{A generic configuration for the $H-S > L$ case.}
    \label{fig6}
\end{figure}

\textbf{\textit{H -- S > L} Case:} In the first case ($H-S>L$), the length of the truncated carry chain is greater than the number of OR gates. A generic configuration for such a case is shown in Figure~\ref{fig6}.
{}
The replacement of FAs with OR gates results in error values between $0$ to $2^{L}-1$, and the carry-chain truncation induces an error equal to $2^{H-S}$. 
Hence, the total error of the approximate block ranges from $0$ to $2^{L}-1$ and from $2^{H-S}$ to $2^{H-S}+2^{L}-1$. 
As there is no carry being propagated from the OR gates part to higher bits and the inputs bits are assumed to be independent, the errors generated in the OR gates part can be considered independent of the error generated due to carry chain truncation. 
Therefore, to compute the PMF of error of the complete approximate block, we can simply convolve the PMF of $E_{OR}$ with the PMF of error of the part between the location of carry chain truncation and bit location $L$ (i.e., the central part of the block). 
Using Eq.~\ref{eq9} and the method presented in Section~\ref{sec_carry_trunc} for computing the combined probability of a set of carry generation events, we can compute the PMF of error of the central part in this case using the following equation. 

\begin{equation} 
    Pr(E_{CP}=y)=
    \begin{cases} 
     \sum_{j=2^{H-S-L}}^{2^{H-S-L+1}-2}P_Z(j;H-S-L) & \text{$y=2^{H-S}$} \\
     1 - \sum_{j=2^{H-S-L}}^{2^{H-S-L+1}-2}P_Z(j;H-S-L) & \text{$y=0$} \\
     0 & otherwise
    \end{cases}
\label{eqation15010}
\end{equation} 

Here, $Pr(E_{CP}=y)$ represents the combined probability of all the events in which the error of the central part of the block is $y$ and $Z$ is the sum of $A_{sub} = [a_{H-S-1}, ..., a_{L}]$ and $B_{sub} = [b_{H-S-1}, ..., b_{L}]$. 
Using the above equations, the PMF of error of the complete approximate block can be computed using the following equation. 

\begin{equation}
Pr(E_{Approx\_Blk}) = Pr(E_{OR}) \circledast Pr(E_{CP})
\label{eq:conv_truncate_or}
\end{equation}

\begin{figure}
    \centering
    \includegraphics[scale=0.6]{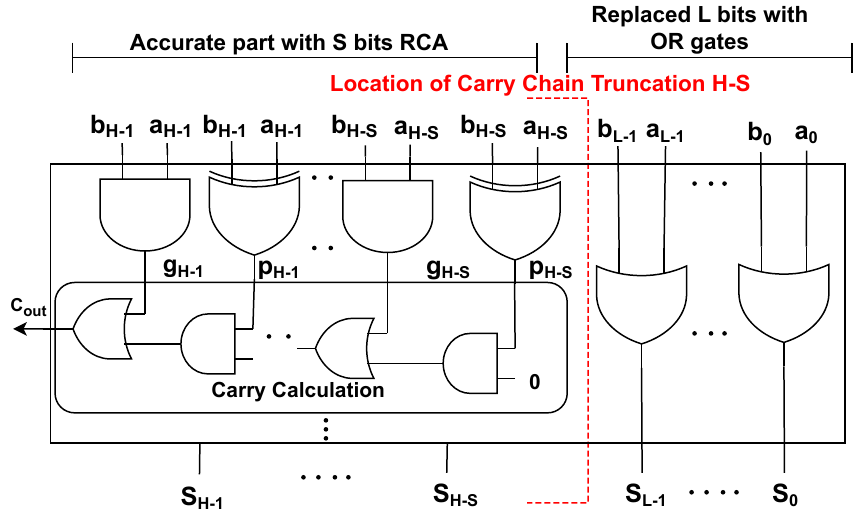}
    \caption{A generic configuration for the $H-S = L$ case.}
    \label{fig7}
\end{figure}

\textbf{\textit{H -- S = L} Case:} In the second case ($H-S=L$), the length of the truncated carry chain is equal to the number of OR gates. 
A generic configuration for such a case is shown in Figure~\ref{fig7}. 
As in this case the location of the carry-chain truncation is the same as the end of the OR gates part, errors in the output of the approximate block are induced only due to the replacement of FAs with OR gates. 
Hence, the PMF of error of the complete approximate block is equivalent to the PMF of $E_{OR}$ of the block, as shown in the following equation. 

\begin{equation}
Pr(E_{Approx\_Blk}) = Pr(E_{OR})
\label{Eq:Case_2_final}
\end{equation}
{}

\begin{figure}[ht]
    \centering
    \includegraphics[scale=0.45]{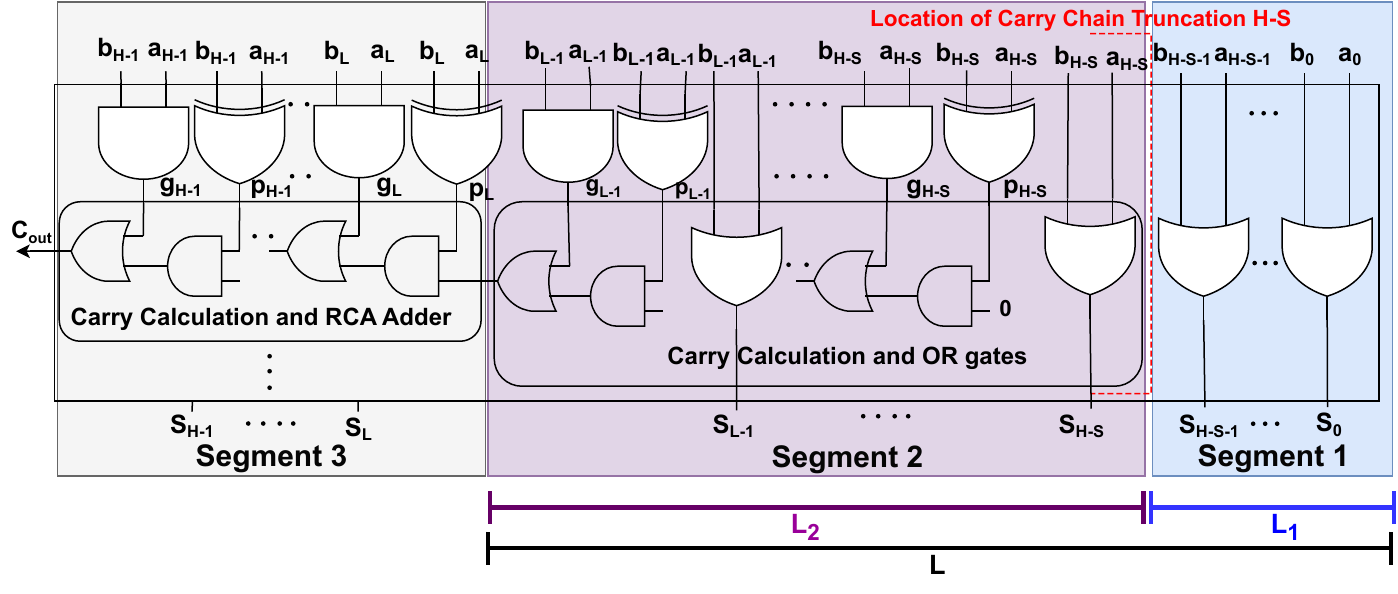}
    \caption{A generic configuration for the $H-S < L$ case. The configuration can be divided into three segments: (1) Sum generation using OR gates with no carry propagation to higher locations; (2) Sum generation using OR Gates with carry propagation to higher locations; and (3) Accurate part of the adder. $L$ is the number of bit locations where the sum is computed using OR gates, $L_1$ is the length of Segment~1, and $L_2$ is the length of Segment~2.}
    \label{fig8}
\end{figure}

\textbf{\textit{H -- S < L} Case:} In the last case ($H-S < L$), the length of the truncated carry chain is smaller than the number of OR gates. A generic configuration for such a case is shown in Figure~\ref{fig8}.
{}

In this case, errors are caused by the computations in the OR gates part and/or truncated carry chain. 
Some inputs can lead to both types of errors. 
Therefore, to compute the PMF of error in this case, we divide the approximate block into multiple segments. 
The first segment is the portion where FAs are approximated with OR gates and there is no carry propagation from the corresponding bits to higher locations, the second segment is the portion where approximate sum generation is performed using OR gates and there is carry propagation to higher locations, and the third segment is the accurate part of the adder. The three segments for an example case are shown in Figure~\ref{fig8}. 

Assuming the bits to be independent, we can compute the PMF of the first segment independently of the second and third segments using Eqs.~\ref{eqsum_error} and~\ref{eq:or_error}. The range of error of this segment is from $0$ to $2^{H-S}-1$. Hence, we can represent the PMF of error of the first segment using the following equation:

\begin{equation}
     Pr(E_{OR}=x)=(\frac{1}{4})^{len(I)}
      . (\frac{3}{4})^{H-S-len(I)} 
\label{eq:or_error_S42}
\end{equation}

where, $0 \leq x \leq 2^{H-S}-1$, $I$ represents the set of bit locations where generate signal is 1 for a given value $x$, and $len(I)$ represents the number of elements in the set $I$. 

\begin{figure}[h]
    \centering
    \includegraphics[width=0.4\linewidth]{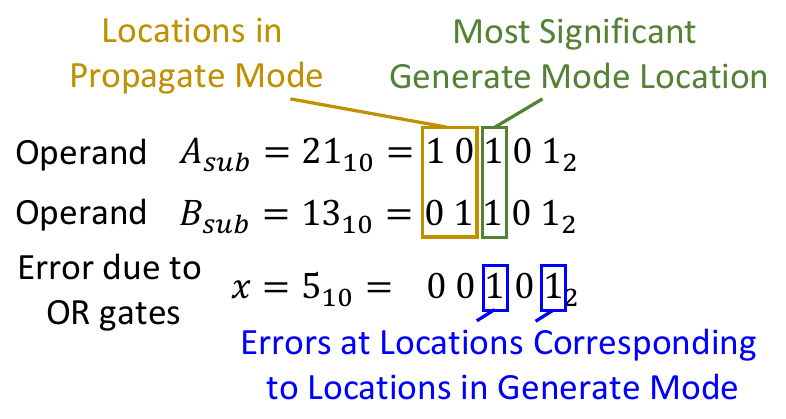}
    \caption{An example of the second segment of an approximate block with carry-out equals 1 in $H - S < L$ case}
    \label{Fig:Case3_example}
\end{figure}

For computing the PMF of error of the second segment, we consider two different cases: one where the carry-out of the segment is $1$ and the other where the carry-out of the segment is $0$. For the case where carry-out equals $1$, assuming the input bits to be independent and uniformly distributed, we can model the probability distribution using the following equation: 

\begin{equation}
     Pr(E=x-2^{L_2})=(\frac{1}{2})^{L_{2_1}}.(\frac{1}{4})^{len(I)}.(\frac{3}{4})^{L_{2_2}-len(I)} 
\label{eq:E_Case3_Cout1}
\end{equation}
Here, $x$ represents the error in the OR gates part (excluding the carry chain circuitry), $I$ represents the set of locations that are in generate mode for the given value of $x$, $L_{2_1}$ is the number of bit locations from MSB of the segment to the most significant location in generate mode (excluding the generate mode location), and $L_{2_2}$ is the number of bit locations from LSB of the segment to the most significant location in generate mode (including the generate mode location). Eq.~\ref{eq:E_Case3_Cout1} is valid only for the cases where $-2^{L_2}+1 \leq E \leq -1$, i.e., for cases where at least one bit location is in generate mode and all the bit locations from the most significant bit location in generate mode until the most significant end of the segment are in propagate mode (including $L_{2_1} =0$ case), see Figure~\ref{Fig:Case3_example} for an example of such a case. 

{}
Similarly, for the case where carry-out equals $0$, we can model the probability distribution using the following equation: 
\begin{equation}
     Pr(E=x)= \frac{\sum_{i=1}^{L_{2_1}}{L_{2_1} \choose i}.2^{L_{2_1} - i}}{4^{L_{2_1}}}.(\frac{1}{4})^{len(I)}.(\frac{3}{4})^{L_{2_2}-len(I)} 
\label{eq:E_Case3_Cout0}
\end{equation}

Eq.~\ref{eq:E_Case3_Cout0} is valid for all the cases where $1 \leq E \leq 2^{L_{2_2}}-1$, i.e., for cases where at least one bit location is in generate mode and at least one bit location in the locations from the most significant bit location in generate mode until the most significant end of the segment is in carry-kill mode. Finally, for $E=0$ case, we can compute the probability using the following equation:

\begin{equation}
     Pr(E=0)= (\frac{3}{4})^{L_2}
\label{eq:E_Case3_0_Error}
\end{equation}

which covers all the cases where there is no generate signal in the bit locations corresponding to the second segment. 
Using the above equations, the probability distribution of the complete approximate block can be computed by first mapping the PMFs to their corresponding error ranges and then convolving the distribution of the first segment with the distribution of the second segment.

\subsection{Evaluation of the PMF of Error of Approximate Blocks with Carry-out set to $0$}

As shown in Figure~\ref{fig4}, the carry-out signal of an approximate block may or may not be connected to the carry-in of the subsequent block, which is mainly based on the location of the approximate block in the adder configuration.
The analytical models presented in the above subsection are mainly designed for approximate blocks whose carry-out is connected to the subsequent block in the adder. 
Therefore, to cover all the possible configurations, there is a need to extend the models for approximate blocks whose carry-out is discarded (i.e., not connected to the subsequent block in the adder). 
To achieve this, we define an approximate HA ($HA_{Approx.}$) and an approximate FA ($FA_{Approx.}$) with carry-out set to 0. 
The truth tables of both are presented in Tables~\ref{tab:1-bit_EV_withoutCin} and~\ref{tab:1-bit_EV_withCin}, respectively. 
The approximate HA design is for the MSB location for the cases where $S=1$, and the approximate FA design is for the cases where $S>1$. 
Assuming the inputs to be uniformly distributed, the PMFs of these approximate HA and approximate FA designs can be represented using Eqs.~\ref{eq:1MSBwithoutCin} and~\ref{eq:1MSBwithCin}, respectively. 

\begin{table}[h]
  \centering
  \caption{Truth table of approximate HA ($HA_{Approx.}$). The error cases are marked in red.}
   \adjustbox{max width=\columnwidth}{%
    \begin{tabular}{|c|c|c|c|c|}
    \hline
    a & b & $C_{out}$ & Sum & Error Value \\ \hline
    0 & 0 & 0 &	0&	0 \\\hline
      0 & 1 & 0 &	1&	0 \\\hline
      1 & 0 & 0 &	1&	0 \\\hline
      1 & 1 & \textcolor{red}{0} &	0&	\textcolor{red}{2} \\\hline
    \end{tabular}}%
  \label{tab:1-bit_EV_withoutCin}%
\end{table}

\begin{table}[h]
  \centering
  \caption{Truth table of approximate FA ($FA_{Approx.}$). The error cases are marked in red.}
   \adjustbox{max width=\columnwidth}{%
    \begin{tabular}{|c|c|c|c|c|c|}
    \hline
    a & b & $C_{in}$ & $C_{out}$ & Sum & Error Value \\ \hline
      0 & 0 & 0 & 0 &	0&	0 \\\hline
      0 & 1 & 0 & 0 &	1&	0 \\\hline
      1 & 0 & 0 & 0 &	1&	0 \\\hline
      1 & 1 & 0 & \textcolor{red}{0} &	0&	\textcolor{red}{2} \\\hline
      0 & 0 & 1 & 0 &	1&	0 \\\hline
      0 & 1 & 1 & \textcolor{red}{0} &	0&	\textcolor{red}{2} \\\hline
      1 & 0 & 1 & \textcolor{red}{0} &	0&	\textcolor{red}{2} \\\hline
      1 & 1 & 1 & \textcolor{red}{0} &	1&	\textcolor{red}{2} \\\hline
    \end{tabular}}%
  \label{tab:1-bit_EV_withCin}%
\end{table}

\begin{equation}
        Pr(E_{HA_{Approx.}})=
\begin{cases}
\frac{3}{4}& \text{$x=0$} \\
\frac{1}{4} & \text{$x=2^H$} \\
\end{cases}
\label{eq:1MSBwithoutCin}
\end{equation}

\begin{equation}
        Pr(E_{FA_{Approx.}})=
\begin{cases}
\frac{1}{2}& \text{$x=0$} \\
\frac{1}{2} & \text{$x=2^H$} \\
\end{cases}
\label{eq:1MSBwithCin}
\end{equation}

As input bits are assumed to be independent of each other, the replacement of the MSB location FA of an approximate block with $HA_{Approx.}$ or $FA_{Approx.}$ can be modeled using the following equation. 

\begin{equation}
    Pr(E_{Approx\_Blk\_C_{out}=0}) =
    Pr(E_{Approx\_Blk}) \circledast
    Pr(E_{AU_{Approx.}})
    \label{eq34}
\end{equation}

Here, $Pr(E_{Approx\_Blk})$ represents the PMF of error of the approximate block from Section~\ref{sec:evaluationapp} considering carry-out signal is propagated to the subsequent block, $Pr(E_{AU_{Approx.}})$ represents the PMF of error of $HA_{Approx.}$ or $FA_{Approx.}$ based on the configuration of the approximate block, and $Pr(E_{Approx\_Blk\_C_{out}=0})$ represents the PMF of error of the complete approximate block considering carry-out signal is set to $0$. 

\subsection{Evaluation of the PMF of Error of an HBAA Configuration}
\label{sec_approx_block_total}
The $N$-bit HBAA is divided into $k$ blocks, each having $H$-bit length. 
$l$ blocks are heterogeneous approximate blocks, and the rest are accurate blocks. 
The sources of error in the output are the errors in the associated approximate blocks. 
As the blocks are independent of each other, the PMF of error value across the HBAA can be calculated by the convolution of the PMFs of all the heterogeneous approximate blocks. 
Thus, the PMF of error value of the approximate adder ($Pr_{EV\_A}$) can be written as:
\begin{equation}
    Pr(E_{Approx\_HBAA})=
    Pr(E_{Blk\_1}) \circledast
    Pr(E_{Blk\_2})
    \circledast ... \circledast Pr(E_{Blk\_l})
    \label{eq34}
\end{equation}
Where $Pr(E_{Blk\_l})$ is the PMF of error of the most significant (i.e., $l^{th}$) approximate block.

\subsection{Evaluation of HBAA's MED and ER}
\label{MED_evaluation}
Mean Error Distance (MED) is considered an important criterion to compare approximate adders. MED can be calculated using the PMF of error by taking the weighted average of all error distances. 
Hence, it is calculated using Eq.~\ref{MED_equation}.
\begin{equation}
    MED=E[ED]=\sum_{i=-\infty}^{\infty}|i| PMF(i)
    \label{MED_equation}
\end{equation}
where $PMF$ is the PMF of error of the approximate adder (in our case, HBAA), and $PMF(i)$ corresponds to the probability of error value equals $i$. Moreover, the Error Rate (ER) can be obtained by adding the probabilities of all non-zero error values from the PMF of error evaluated by our proposed analytical model. 
\color{black}

\section{Analytical Modeling for Estimating Hardware Metrics of HBAA designs}
\label{Sec:hardware_metric_models}
In real-world error-resilient applications, an acceptable accuracy level, which is identified by ED, ER, or MED, must be satisfied. 
Therefore, it is important to effectively use the available error budget for improving the efficiency of the underlying hardware/system. 
Metrics like area, delay, and power are commonly used to estimate the performance and efficiency of the hardware. 
Configurations that offer the best accuracy-efficiency trade-offs are identified by exploring the complete design space using both error and performance metrics.  
Hence, alongside error estimation models, performance estimation models are also required. 
In this work, we extend the estimation method proposed in~\cite{dutt2018analysis} to build models for computing the hardware metrics of HBAA configurations.

\begin{figure}
    \centering
    \includegraphics[width=0.5\columnwidth]{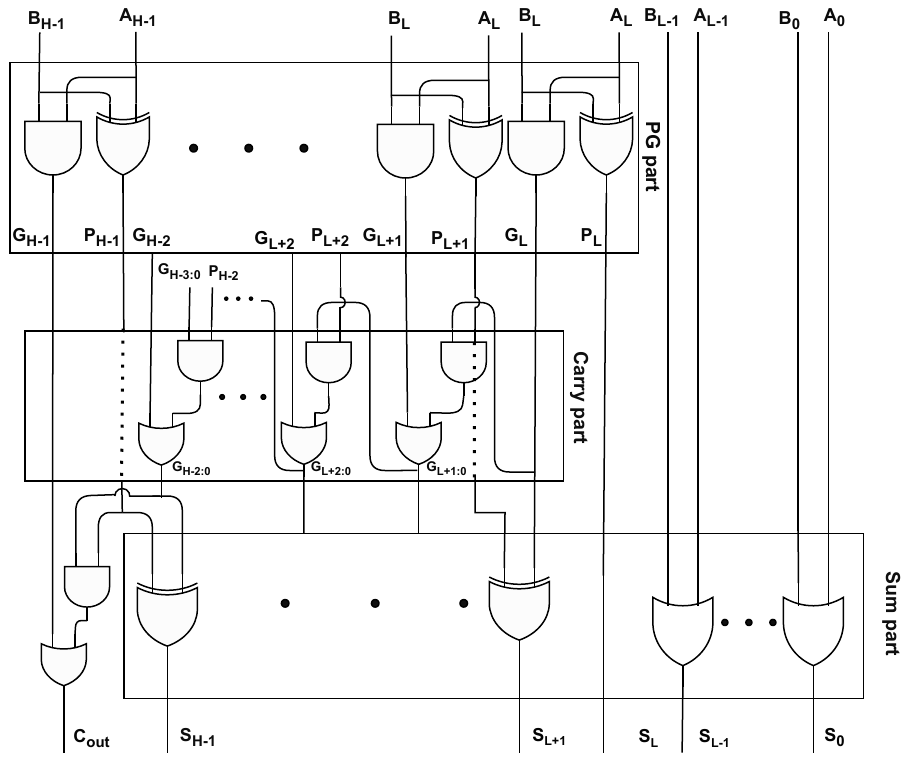}
    \caption{Gate-level implementation of an approximate block of HBAA.}
    \label{PG_block}
\end{figure}

Conventional adders such as RCA are composed of three main parts, i.e., Propagate and Generate (PG) signal generation part, carry generation part, and sum computation part~\cite{dutt2018analysis}\cite{weste2010cmos}. 
Any abstraction level of a digital design, from the highest behavioral level to the lowest device level, can be considered to estimate its performance/hardware metrics. 
In this work, the gate-level abstraction is considered for modeling the hardware characteristics of adder designs. 
We consider 2-input gates, e.g., AND, OR, NAND, and NOR, as the elementary gates for implementing adder designs. 
Other gates, such as XOR and XNOR, can be expressed in terms of the above-mentioned elementary gates. 
We neglect NOT (inverter) gates in the delay and area estimation. 
Thus, in this work, a circuit is modeled by 2-input gates, and gate-level depth and gate count are used to estimate delay and area, respectively. 
In our estimation model, the XOR gate is constructed from three 2-input gates, i.e., two 2-input AND and one 2-input OR. 
Thus, the gate-level depth and gate count of the XOR gate are 2 and 3, respectively. 
The gate-level implementation of an approximate block of the proposed HBAA is shown in Figure~\ref{PG_block}, which is used in this work to compute the gate-level depth and gate count of HBAA configurations.
{}
\medskip
\begin{figure}
    \centering
    \includegraphics[scale=0.45]{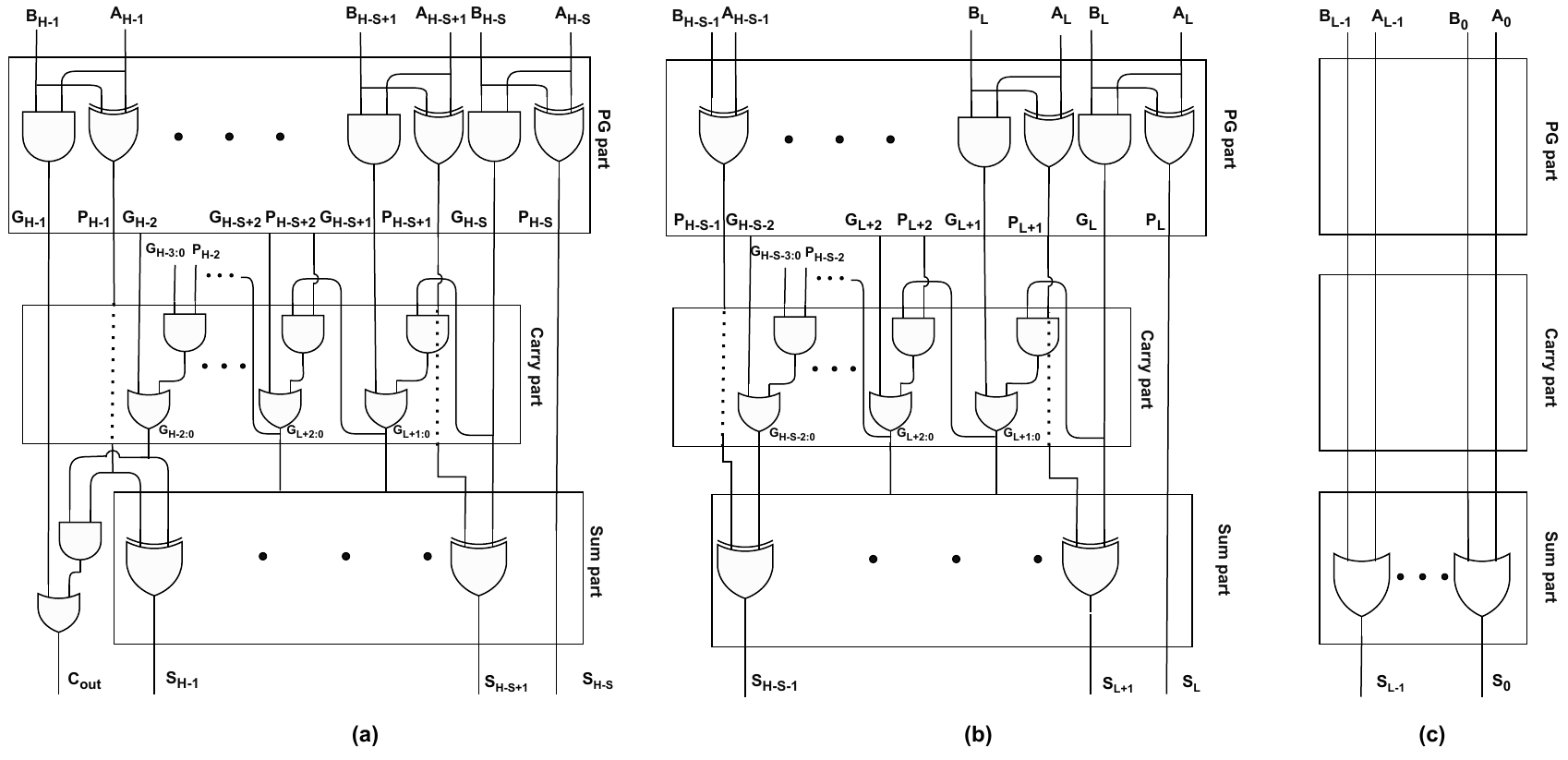}
    \caption{Gate-level implementation of different segments of an approximate block with $H-S > L$ configuration.}
    \label{Fig:Individual_Gate_Level_Implementations_of_Segments}
\end{figure}{}
\begin{figure}
    \centering
    \includegraphics[scale=0.45]{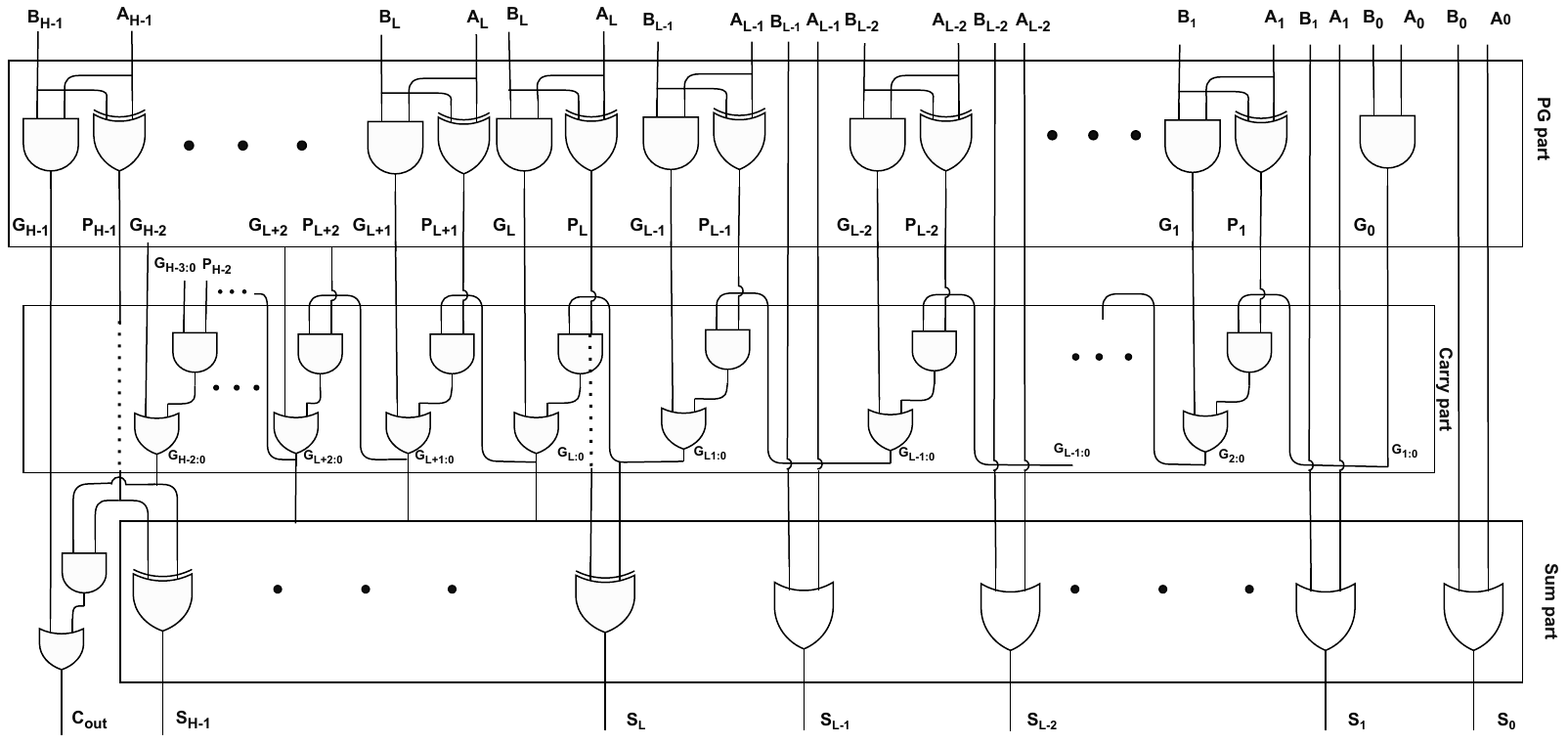}
    \caption{Gate-level implementation of the most significant segment (apart from the truncated OR gates part) of an approximate block with $H-S < L$ configuration.}
    \label{Fig:Individual_Gate_Level_Implementations_of_Segments_2}
\end{figure}{}
\subsection{Delay Estimation}
As shown in Figure~\ref{PG_block}, the gate-level implementation of an approximate block of HBAA consists of three parts. 
The length of each part depends on the configuration of the approximate block, i.e., on $H$, $L$, and $S$ values of the block. 
To construct a model for delay estimation, we first consider the case of $H-S > L$. 
As earlier shown in Figure~\ref{fig6}, in such cases the block configuration can be divided into three independent segments, i.e., OR gates part, the central part, and the accurate most significant part. 
The gate-level implementations of the three parts are shown in Figure~\ref{Fig:Individual_Gate_Level_Implementations_of_Segments}, where Figure~\ref{Fig:Individual_Gate_Level_Implementations_of_Segments}c shows the OR gates part, Figure~\ref{Fig:Individual_Gate_Level_Implementations_of_Segments}b shows the central part, and Figure~\ref{Fig:Individual_Gate_Level_Implementations_of_Segments}a shows the accurate most significant part. 
As there is no carry propagation between these segments, the gate-level depth of each can be computed individually and the maximum of these can be used as the depth estimate for the whole approximate block. 
From Figure~\ref{Fig:Individual_Gate_Level_Implementations_of_Segments}c, we observe that the depth of the OR gates part is 1. 
From Figure~\ref{Fig:Individual_Gate_Level_Implementations_of_Segments}b, we observe that the depth of the central part depends on the length and depth of the PG, Carry, and Sum parts of the gate-level implementation. 
The depth of the PG part is 2 when $H-S-L \geq 1$. 
The depth of the Carry part is 0 when $H-S-L \leq 2$ while it is $2(H-S-L-2)$ when $H-S-L > 2$. 
And, the depth of the Sum part is 0 when $H-S-L \leq 1$ while it is 2 when $H-S-L \geq 2$. 
Thus, the depth of the central part can be summarized using the following equation. 

\begin{equation}
    Gate\_Depth=
2(H-S-L) \quad\quad\quad\quad \text{when}\;\text{$H-S-L \geq 0$} 
\label{Eq:Central_Part_Depth}
\end{equation}

Similar to the case of the central part, from Figure~\ref{Fig:Individual_Gate_Level_Implementations_of_Segments}a, we observe that the depth of the accurate most significant part depends on the length and depth of the PG, Carry and Sum parts of the gate-level implementation. 
The depth of the PG part is 2 when $S \geq 1$. 
The depth of the Carry part is 0 when $S \leq 2$ while it is $2(S-2)$ when $S > 2$. 
And, the depth of the Sum part is 0 when $S \leq 1$ while it is 2 when $S \geq 2$. 
Thus, the depth of the accurate most significant part can be summarized using the following equation. 

\begin{equation}
    Gate\_Depth=
2S \quad\quad\quad\quad \text{when}\;\text{$S \geq 0$} 
\label{Eq:MS_Accurate_Part_Depth}
\end{equation}

Using the depths of all the parts shown in Figure~\ref{Fig:Individual_Gate_Level_Implementations_of_Segments}, the delay of an approximate block with $H-S>L$ can be summarized as: 

\begin{equation}
    Delay_{Approx\_Block}=
max(2C_dS, 2C_d(H-S-L)) 
\label{Eq:H-S>L_Delay}
\end{equation}
Where $C_d$ is a technology dependent constant for delay. 

For the $H-S=L$ case, as only the OR gates part and the accurate most significant part can be present, the delay of such an approximate block can be computed using the following equation.

\begin{equation}
    Delay_{Approx\_Block}=
max(2C_dS, C_d) 
\label{Eq:H-S=L_Delay}
\end{equation}

For the $H-S < L$ case, we define another type of segment shown in Figure~\ref{Fig:Individual_Gate_Level_Implementations_of_Segments_2}. The gate-level depth of this segment can be computed using the following equation. 

\begin{equation}
    Gate\_Depth=
2S \quad\quad\quad\quad \text{when}\;\text{$S \geq 0$} 
\label{Eq:MS_Accurate_Part2_Depth}
\end{equation}

Note that even when $L=H$, the above equation is valid, as there are two additional gates installed in parallel to the sum part for generating the carry-out signal. 
Hence, the delay of an approximate block with $H-S < L$ can be computed using the following equation.

\begin{equation}
    Delay_{Approx\_Block} = 2C_dS
\label{Eq:H-S<L_Delay}
\end{equation}

Using the above equations, we can generalize the delay of an approximate block using the following equation.

\begin{equation}
\footnotesize
    Delay_{Approx\_Block} = 
    \begin{cases}
    max(2C_dS, 2C_d(H-S-L)) & H-S > L \\
    max(2C_dS, C_d) & H-S=L \\
    2C_dS & H-S < L
    \end{cases}
\label{Eq:Complete_Approx_Block_Delay}
\end{equation}

The above equation is for the case where the carry-out signal of the block is propagated to the next block. However, if the carry-out signal is not propagated, Eq.~\ref{Eq:Complete_Approx_Block_Delay} changes to the following equation because of the absence of the last two gates used for carry-out signal generation in Figure~\ref{Fig:Individual_Gate_Level_Implementations_of_Segments_2}. 

\begin{equation}
\footnotesize
    Delay_{Approx\_Block} = 
    \begin{cases}
    max(2C_dS, 2C_d(H-S-L)) & H-S > L \\
    max(2C_dS, C_d) & H-S=L \\
    2C_dS & H-S < L \text{ and } L \neq H \\
    C_d & H-S < L \text{ and } L = H
    \end{cases}
\label{Eq:Complete_Approx_Block_Delay_0}
\end{equation}

Besides a model for estimating the delay of approximate blocks, we need a model for computing the delay of the accurate blocks used in the most significant part of HBAA configurations. 
Using the method proposed in~\cite{dutt2018analysis}, the delay of an $H$-bit accurate block can be computed using the following equation. 

\begin{equation}
    Delay_{Accurate\_Block}=2C_d(H+1) 
\label{delay_equation1}
\end{equation}

Using the delays of individual approximate and accurate blocks, the delay of an HBAA configuration can be computed by using the following equation.

\begin{equation}
    Delay_{HBAA\_Config.}=
    \begin{cases}
    max(C_dS_j + \sum_{i=j+1}^{k} Delay_{Block\_i}, Delay_{Block\_{j}}, ..., Delay_{Block\_1}) & S_j\leq1 \\\
    max(2C_dS_j + \sum_{i=j+1}^{k} Delay_{Block\_i}, Delay_{Block\_{j}}, ..., Delay_{Block\_1}) & S_j>1
    \end{cases}   
\end{equation}

Where $Block\_{j}$ is the most significant approximate block in the given HBAA configuration and $S_j$ represents the carry-chain length to generate the carry-out signal of the $j^{th}$ approximate block.

\subsection{Area Estimation}
The area estimate of an HBAA configuration is calculated based on its gate count. 
As shown in Figure~\ref{PG_block}, a sub-adder block in HBAA is composed of three different parts, i.e., PG, Sum part, and Carry part. 
Therefore, for estimating the area of a sub-adder block of an HBAA configuration, we compute the gate count in each individual part of the block and then sum them up to get the final gate count for the block.  
In an $H$-bit accurate block, the gate count of the PG part is $4H$, the gate count of the sum part is $3H$, and the gate count of the carry part is $2H$. Thus, the overall gate count of an $H$-bit accurate block ($Gate\_Count_{Accurate}$) can be obtained by using the following equation. 

\begin{equation}
    Gate\_Count_{Accurate}= 9H
\label{area_equation}
\end{equation}

The gate count of each part of an approximate block of an HBAA configuration can be computed using the equations mentioned in Table~\ref{tab:table2}.

\begin{table}[ht]
\centering

\caption{Approximate block's gate count }
\adjustbox{max width=\textwidth}{%
\begin{tabular}{|c|c|c|}
\hline
Parts      & Gate count of  Approximate block with Carry-out  & Gate count of Approximate block with Carry-out set to $0$ \\\hline
PG part    & 
    $Gate\_Count_{PG}=
    \begin{cases}
4(H-L)-1 & \text{$H-S > L$} \\
4S & \text{$H-S=L$} \\
4S-3 & \text{$H-S<L$} \\
\end{cases}$ &
$Gate\_Count_{PG}=
    \begin{cases}
4(H-L)-2 & \text{$H-S > L$} \\
4S-1 & \text{$H-S=L$} \\
4S-4 & \text{$H-S<L$} \\
\end{cases}$
        \\\hline
Sum part   & 
    $Gate\_Count_{Sum}=
    \begin{cases}
    3(H-L-2)+L & \text{$H-S > L$} \\
    3(H-L-1)+L & \text{$H-S = L$} \\
    3(H-L)+L & \text{$H-S < L$} \\
    \end{cases}$ &
    $Gate\_Count_{Sum}=
    \begin{cases}
    3(H-L-2)+L & \text{$H-S > L$} \\
    3(H-L-1)+L & \text{$H-S = L$} \\
    3(H-L)+L & \text{$H-S < L$} \\
    \end{cases}$ 
                   \\\hline
Carry part & 
  $Gate\_Count_{Carry}=
    \begin{cases}
0 & \text{$H-S \geq L$  and $S \leq 1$ and $H-S-L\leq 2$}  \\
2(H-L-3) & \text{$H-S > L$  and $S > 1$ and $H-S-L > 2$} \\
2(S-1) & \text{$H-S \leq L$ and $S > 0$} \\
\end{cases}$ &  
  $Gate\_Count_{Carry}=
    \begin{cases}
0 & \text{$H-S > L$  and $S \leq 2$ and $H-S-L\leq 2$}  \\
2(H-L-4) & \text{$H-S > L$ and $S > 2$ and $H-S-L > 2$} \\
0 & \text{$H-S \leq L$ and $S \leq 1$} \\
2(S-2) & \text{$H-S \leq L$ and $S > 1$} \\
\end{cases}$
      \\ \hline  
$Gate\_Count_{Approximate}$ &  \multicolumn{2}{|c|}{$Gate\_Count_{PG}+Gate\_Count_{Sum}+Gate\_Count_{Carry}$}\\\hline 
\end{tabular}}
\label{tab:table2}
\end{table}

$Gate\_count_{Approximate}$ presents the gate count of an approximate block. The area estimate of an $N$-bit HBAA is equivalent to the sum of the areas of all the accurate and approximate blocks in the adder. 
Thus, the area estimate of an HBAA configuration can be computed using the following equation. 

\begin{equation}
    Area_{HBAA\_Config.}=C_a(\sum_{i=1}^{k} Gate\_Count_{Block\_i})
\label{area_equation_tottal}
\end{equation}

Where $Gate\_Count_{Block\_i}$ represents the gate count of the $i^{th}$ block in the configuration and $C_a$ is a technology dependent constant for area. 

\subsection{Power Estimation}
Power consumption of a  digital circuit is estimated based on the following two components:
 \begin{itemize}
 \item \textbf{Dynamic Power:} The dynamic power consumption ($P_d$) of a digital circuit is directly proportional to its area and delay if the clock frequency is assumed to be fixed~\cite{dutt2018analysis}. Thus, $P_d$ of an HBAA configuration at a fixed clock frequency can be estimated by using Eq.~\ref{dynamic_power}.
\begin{equation}
    P_d\approxprop (area . delay) \Rightarrow P_d=C_{pd}(Gate\_Count_{HBAA\_Config.} . Gate\_Depth_{HBAA\_Config.})
\label{dynamic_power}
\end{equation}
Where $C_{pd}$ is a technology dependent constant for dynamic power.
 \item \textbf{Static Power:} According to \cite{dutt2018analysis}, the static power consumption ($P_s$) of a digital circuit is directly proportional to its area. Thus, $P_s$ of an HBAA configuration can be estimated using the following equation. 
\begin{equation}
    P_s\approxprop area \Rightarrow P_s=C_{Ps}(Gate\_Count_{HBAA\_Config.})
\label{static_power}
\end{equation}
Where $C_{Ps}$ is a technology dependent constant for static power.
 \end{itemize}
 
 As a result, the total power consumption can be estimated by the sum of dynamic and static power consumption. 
 \begin{equation}
    P = P_s + P_d
\label{power_equation}
\end{equation}
Similar to~\cite{dutt2018analysis}, we obtained the technology dependent delay, area and power constants by implementing a 2-input NAND gate and extracting its hardware characteristics. 
We synthesize a 2-input NAND gate using \emph{Synopsys Design Compiler} with the Nangate $15nm$ FinFET Open Cell Library. 
For the power constant, similar to~\cite{dutt2018analysis}, we compute $C_p$, which is equivalent to $C_{Pd}+C_{Ps}$.    
The values of the constants derived from the implementation of a 2-input NAND gate using the Nangate $15nm$ FinFET Open Cell Library are presented in Table~\ref{tab:table3}. 
\begin{table}[ht]
\centering
\caption{Constant Factor of 15 $nm$ technology }
\adjustbox{max width=\textwidth}{%
\begin{tabular}{|c|c|}
\hline
Constant Factor & Value \\\hline
$C_d$    & 1.26 $ps$         \\\hline
$C_a$   & 0.14 $\mu m^2$       \\\hline
$C_p$ & 1.74 $\mu W$      \\ \hline  
\end{tabular}}
\label{tab:table3}
\end{table}
\color{black}

\section{Results and Discussion}
\label{Sec:Results}

In this section, we compare the design space of our proposed HBAA with that of different state-of-the-art approximate adder designs in order to highlight the significance of HBAA for providing better accuracy-efficiency trade-offs. 
We also discuss the accuracy of our proposed analytical model for computing the error metrics of HBAA configurations. 

\subsection{Error Metrics}
In this work, we used (1) Mean Error Distance (MED)~\cite{6569370, shafique2015low, 6226361, 6797866, mazahir2016probabilistic}, (2) Normalized Mean Error Distance (NMED)~\cite{shafique2015low, 6226361}, and (3) Error Rate (ER)~\cite{ayub2017statistical} as the error metrics for comparing different approximate adders. The definitions of these error metrics are presented below. 

\textbf{Mean Error Distance (MED)} of an n-bit approximate adder is defined as: 

\begin{equation}
MED = \frac{1}{2^{2n}}\sum_{i=0}^{2^{n}-1}\sum_{j=0}^{2^{n}-1}|S_{accu}(i,j) - S_{approx}(i,j)|
\label{eq-MED}    
\end{equation}

Here, $S_{accu}(i,j)$ defines the accurate sum of $i$ and $j$ while $S_{approx}(i,j)$ defines the approximate sum of $i$ and $j$ (computed using the given approximate adder).  

\textbf{Normalized Mean Error Distance (NMED)} of an n-bit approximate adder is defined as: 
\begin{equation}
 NMED = \frac{MED}{2^n} = \frac{1}{2^n}(\frac{1}{2^{2n}}\sum_{i=0}^{2^{n}-1}\sum_{j=0}^{2^{n}-1}|S_{accu}(i,j) - S_{approx}(i,j)|)
\label{eq-NED}    
\end{equation}

\textbf{Error Rate (ER)} of an n-bit approximate adder is defined as the percentage of erroneous outputs among all outputs and is computed using: 
\begin{equation}
    ER = \frac{1}{2^{2n}}\sum_{i=0}^{2^{n}-1}\sum_{j=0}^{2^{n}-1} f(|S_{accu}(i,j) - S_{approx}(i,j)|) 
    \label{eq-ER}  
\end{equation}

where

\begin{equation}
     f(x)=
    \begin{cases}
1 & x\neq0 \\
 0 & x=0 \\
\end{cases}
\label{f_equation}
\end{equation}

\subsection{Accuracy of the Proposed Analytical Model for Computing Error Metrics}

In this section, we evaluate the accuracy of our proposed analytical model for computing the error metrics of HBAA configurations. 
To achieve this, we compare the results generated using the proposed analytical model with the results generated using Monte Carlo simulation. 
Table~\ref{tab:error_16} presents the MED values computed using the proposed analytical model and Monte Carlo simulations for different randomly selected 16-bit HBAA configurations. 
The table also presents the accuracy of the values computed using the analytical model by comparing them with the results generated using Monte Carlo simulations. 
The results show that the proposed analytical model is capable of generating error metrics fairly close to that of Monte Carlo simulations, i.e., on average 99.64\% accuracy.
Note that for this analysis, each Monte Carlo simulation result is computed using 10~million randomly generated input combinations. 

\begin{table}[ht]
  \centering
  \caption{Accuracy of our proposed analytical model for computing MED of 16-bit HBAA configurations}
   \adjustbox{max width=\columnwidth}{%
    \begin{tabular}{|c|c|>{\centering\arraybackslash}m{3cm}|>{\centering\arraybackslash}m{3cm}|>{\centering\arraybackslash}m{3cm}|}
    \hline
    16-bit Adder & Configuration & MED calculated using Monte Carlo Simulation &  MED calculated using Analytical Model & Accuracy of the Analytical Model \\ \hline
    \multirow{15}{*}{HBAA} 
    & \{1,4,2,3\}\{4,0,3,1\} & 9310.41 & 9313.64 & 99.97\% \\\cline{2-5}
      & \{4,2\}\{0,2\} & 19.26 &	19.5 & 98.77\% \\\cline{2-5}
      &\{2,2,2,1\}\{0,0,0,1\} & 47.28 & 47.5 &	99.54\% \\\cline{2-5}
        &\{6\}\{3\} & 0.25 &	0.25 &	100\% \\\cline{2-5}
        &\{2,1,2,1,0,2,1,1\}\{1,1,2,1,2,1,2,2\} &9491.73 &	9404.39&	99.07\% \\\cline{2-5}
        &\{3,2,3\}\{2,2,1\} &595.41	& 595.85	&99.93\%\\\cline{2-5}
        &\{4,4\}\{0,2\} & 65.45&	66.69	&98.14\%\\\cline{2-5}
        &\{5,2\}\{4,4\}  & 1854.24	&1856.83&	99.86\%\\\cline{2-5}
        &\{2,1,2,1,0,2\}\{0,0,0,1,2,1\} & 1038.74	&1030.72&	99.22\% \\\cline{2-5}
        &\{2,2,2,2,1,2,2\}\{1,0,2,1,2,2,1\}	&3817.06	&3855.7&	99.00\% \\\cline{2-5}
        &\{2,3\}\{0,4\} & 35.88	&35.94	&99.83\% \\\cline{2-5}
        &\{1,1,2,0,0,2\}\{0,1,1,1,2,0\}	&1519.34	&1523.46	&99.73\%\\\cline{2-5}
        &\{4,4\}\{2,3\}	& 72.02	&72.16	&99.81\% \\\cline{2-5}
        &\{2,1,2,1,0,2\}\{1,1,0,1,2,1\} & 1024.68	&1029.76	&99.51\%\\\cline{2-5}
        &\{2,1,2,1,2,2,1,1\}\{1,2,2,1,2,0,2,2\} &9463.51&	9515.9&	99.45\%\\\hline
    \end{tabular}}%
  \label{tab:error_16}%
\end{table}

\begin{figure}[ht]
    \subfloat[]{
        \includegraphics[scale=0.15]{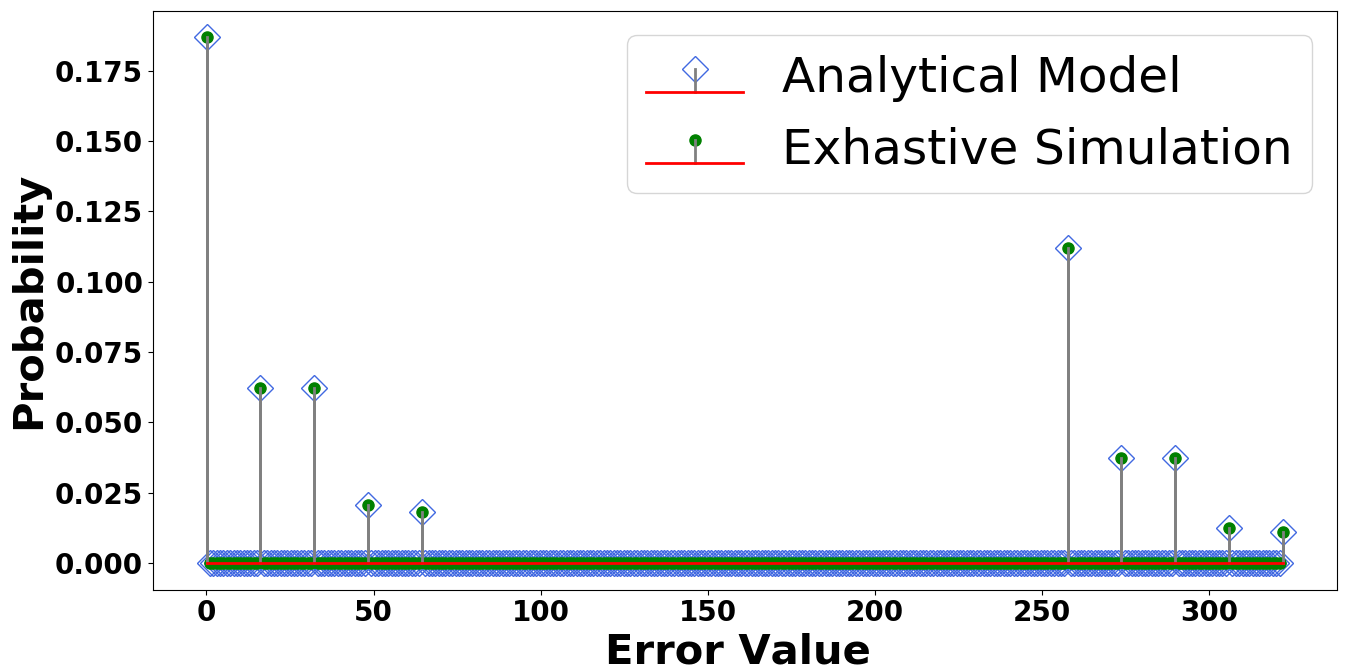}
    }
    \subfloat[]{
         \includegraphics[scale=0.15]{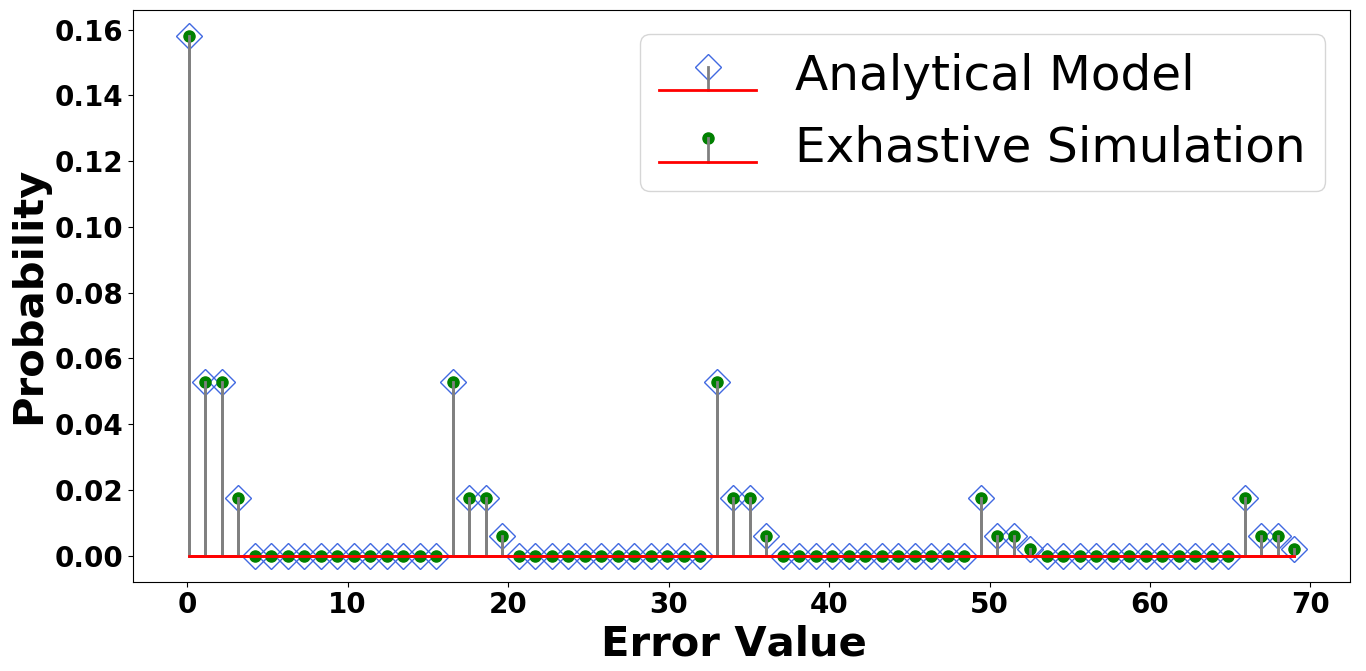}
    }\\
    \subfloat[]{
         \centering
         \includegraphics[scale=0.15]{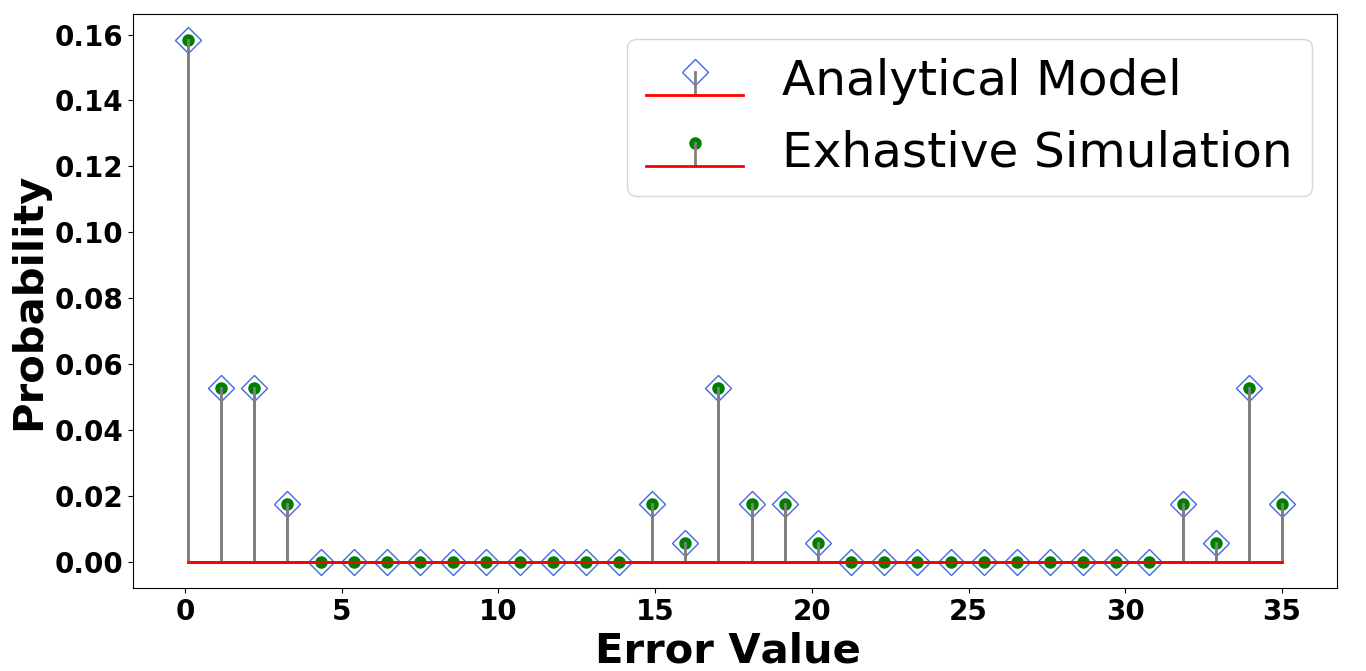}
    }
    \subfloat[]{
         \centering
         \includegraphics[scale=0.155]{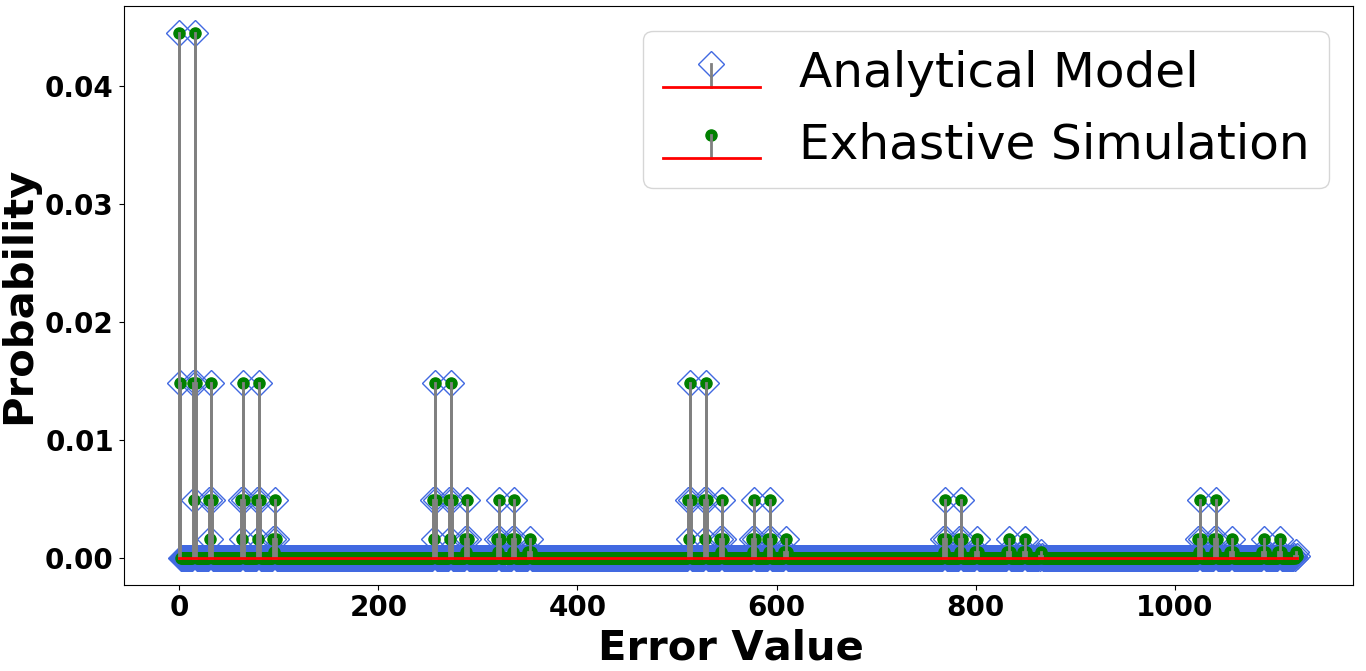}
    }
        \caption{Comparison of the proposed analytical model and exhaustive simulations for generating PMF of error values for 4 different 16-bit HBAA configurations which have 4-bit sub-blocks ($H=4$): (a) HBAA\{[2,2],[0,0]\} (b) HBAA\{[2,2],[2,2]\} (c) HBAA\{[2,2],[3,3]\} (d) HBAA\{[2,1,2],[3,2,2]\}. The results are generated assuming uniform input distribution.}
        \label{figure_pmf}
\end{figure}

To highlight the accuracy of the proposed analytical model for computing the PMF of error values, Figure~\ref{figure_pmf} presents a comparison between PMF of error values generated using the proposed analytical model and PMF of error values generated using exhaustive simulation for 4 different 16-bit HBAA configurations.
The configurations are composed of 4-bit sub-adder blocks, i.e., for all the configurations $H=4$. 
The PMFs shown in Figure~\ref{figure_pmf} are composed of discrete impulses, where each impulse defines the probability of the corresponding error value. 
The figure shows that for each configuration, the PMF generated using the proposed analytical model is exactly the same as the PMF generated using exhaustive simulations. 
Therefore, it can be concluded that the proposed analytical model is capable of providing accurate error estimates.
  
{}

Similar to Table~\ref{tab:error_16}, Table~\ref{tab:error_32} presents MED values computed using the proposed analytical model and Monte Carlo simulations for 6 different randomly selected 32-bit HBAA configurations. 
For this analysis, we performed Monte Carlo simulations using 10~million randomly generated input combinations as well as using 1~billion randomly generated input combinations. 
The table also presents the accuracy of the proposed analytical model in comparison to the 1 billion combinations based Monte Carlo simulations. 
The results show that for all the presented configurations the analytical model generates fairly accurate error estimates, i.e., on average 99.52\% accurate.
The table also highlights that the results generate using 10~million combinations based Monte Carlo simulations and the results generated using 1~billion combinations based Monte Carlo simulations are approximately the same, and therefore, Monte Carlo simulation using 10~million randomly selected input combinations can be used, as they generate good-enough estimates and require just 4.2 minutes per configuration to complete compared to around 6 hours for 1 billion combinations based simulations. 

\begin{table}[ht]
  \centering
  \caption{Accuracy of our proposed analytical model for computing MED of 32-bit HBAA configurations}
   \adjustbox{max width=\textwidth}{%
    \begin{tabular}{|c|c| >{\centering\arraybackslash}m{3cm} |>{\centering\arraybackslash}m{3cm} |c|>{\centering\arraybackslash}m{3cm} |}
    \hline
    32-bit Adder & Configuration & MED computed using Monte Carlo simulation with 10 million combinations & MED computed using Monte Carlo simulation with 1 billion combinations & Analytical Model & Accuracy of Monte Carlo simulation with 1 billion combinations results compared to the analytical model results
 \\ \hline
    \multirow{6}{*}{HBAA}   & \{[4,4,4,2][0,0,0,2]\} & 4095.66  & 4095.59 & 4095.75 &99.99\%  \\\cline{2-6}
       & \{[4,2][0,2]\} & 15.75 & 15.75 & 15.75 &100\%  \\\cline{2-6}
       & \{[4,1][0,3]\} & 7.7543  & 7.75 & 7.75 &100\%  \\\cline{2-6}
      & \{[4,4,4,1][0,0,0,3]\} & 2048.23 & 2047.78 & 2047.75 & 99.99\% \\\cline{2-6}
      & \{[4,4,4,2][0,0,0,3]\} & 3071.39 & 3071.98 & 3071.875 & 99.99\% \\\cline{2-6}
     & \{[4,4,4,4,1][0,0,0,0,3]\} & 32750.77&32766.54&31867.75&97.18\%\\\hline
    \end{tabular}}%
  \label{tab:error_32}%
\end{table}

\subsection{Hardware Metrics and Accuracy of Proposed Hardware Estimation Models}

For hardware metrics, we mainly considered area, delay, and power for comparing different approximate adder configurations. 
We used Verilog HDL to describe our proposed as well as other state-of-the-art approximate adders (i.e., GeAr, SARA, and BCSA~\cite{ebrahimi2019block}). 
To evaluate the accuracy of our proposed hardware estimation models, we synthesized different HBAA, GeAr, SARA, and BCSA configurations using Synopsys Design Compiler and computed their area and delay values.
For synthesis, we used Nangate $15nm$ FinFET Open Cell Library with $0.8V$ operating voltage and $25^oC$ temperature. 
To obtain the power values of adders, we used ModelSim tool to generate the VCD files and then used Synopsys PrimeTime to generate the final power values. 
To generate VCD files, we injected 10~million randomly selected inputs into the netlist of synthesized adders and stored the internal activity information in VCD file format.  

Tables~\ref{tab:hardware_area_16},~\ref{tab:hardware_delay_16}, and~\ref{tab:hardware_power_16} present area, delay and power values for different 16-bit HBAA, GeAr, SARA and BCSA configurations. 
The tables include both the values computed using Synopsys Design Compiler and the values computed using the proposed hardware estimation model. 
The results show that the proposed hardware estimation models offer highly accurate results, i.e., on average 94.29\% accuracy for area estimates, 94.34\% for delay estimates, and 90.70\% for power estimates.
We also performed a similar comparison for 32-bit approximate adder configurations. 
Tables~\ref{tab:hardware_area32},~\ref{tab:hardware_delay32}, and~\ref{tab:hardware_power32} present the area, delay, and power values for different 32-bit HBAA, GeAr, SARA and BCSA configurations. 
The results show that the proposed hardware estimation model offers on average 94.38\% accuracy for area, 92.31\% for delay, and 91.14\% for power estimates.

\begin{table}[ht]
  \centering
  \caption{Accuracy of our proposed estimation model for computing the Area of 16-bit approximate adders}
   \adjustbox{max width=\columnwidth}{%
    \begin{tabular}{|c|c|>{\centering\arraybackslash}m{3cm}|>{\centering\arraybackslash}m{3cm}|>{\centering\arraybackslash}m{3cm}|}
    \hline
    16-bit Adder & Configuration & Area Estimate using the proposed analytical Model ($\mu m^2$) & Area computed using Synopsys Design Compiler ($\mu m^2$) & Accuracy of area estimation model \\ \hline
    \multirow{15}{*}{HBAA} 
    & \{1,4,2,3\}\{4,0,3,1\} & 8.96 & 9.93 & 90.23\% \\\cline{2-5}
      & \{4,2\}\{0,2\} & 13.15 & 12.87 & 97.82\% \\\cline{2-5}
      &\{2,2,2,1\}\{0,0,0,1\} & 12.05 & 11.59 & 96.03\% \\\cline{2-5}
        &\{6\}\{3\} & 13.58 & 14.27 & 95.16\% \\\cline{2-5}
        &\{2,1,2,1,0,2,1,1\}\{1,1,2,1,2,1,2,2\} & 14 & 13.26 & 94.42\% \\\cline{2-5}
        &\{3,2,3\}\{2,2,1\} & 10.8 & 11.37 & 94.99\% \\\cline{2-5}
        &\{4,4\}\{0,2\} & 12.18 & 13.05 & 93.33\% \\\cline{2-5}
        &\{5,2\}\{4,4\}  & 13.16 & 12.34 & 93.35\% \\\cline{2-5}
        &\{2,1,2,1,0,2\}\{0,0,0,1,2,1\} & 11.2 & 10.34 & 91.68\% \\\cline{2-5}
        &\{2,2,2,2,1,2,2\}\{1,0,2,1,2,2,1\}	& 12.32 & 11.62 & 93.98\% \\\cline{2-5}
        &\{2,3\}\{0,4\} & 14.84 & 14.36 & 96.66\% \\\cline{2-5}
        &\{1,1,2,0,0,2\}\{0,1,1,1,2,0\}	& 13.44 & 12.86 & 95.49\% \\\cline{2-5}
        &\{4,4\}\{2,3\}	& 14.54 & 15.28 & 95.16\% \\\cline{2-5}
        &\{2,1,2,1,0,2\}\{1,1,0,1,2,1\} & 12.88 & 12.34 & 95.62\%\\\cline{2-5}
        &\{2,1,2,1,2,2,1,1\}\{1,2,2,1,2,0,2,2\} & 13.44 & 12.27 & 90.46\%\\\hline
     \multirow{3}{*}{SARA}
     &SARA2&	23.52&	20.86&	87.25\% \\\cline{2-5}
    &SARA4&	25.48&	23.12&	89.79\% \\\cline{2-5}
    &SARA8&	22.96&	24.36&	94.25\% \\\hline
    \multirow{3}{*}{GeAr}  
    & \{16,4,0\}	&17.36&	16.23&	93.04\%\\\cline{2-5}
    &\{16,8,0\}&	19.32&	17.5&	89.60\%\\\cline{2-5}
    &\{16,6,4\}&	25.2&	24.42&	96.81\%\\\hline   
    \multirow{3}{*}{BCSA} 
    &BCSA2&	22.82&	20.56&	89.01\%\\\cline{2-5}
	&BCSA4&	21.7&	22.15&	97.97\%\\\cline{2-5}
	&BCSA8&	21.14&	23.43&	90.23\% \\\hline
    \end{tabular}}%
  \label{tab:hardware_area_16}%
\end{table}
\begin{table}[ht]
  \centering
  \caption{Accuracy of our proposed estimation model for computing the Delay of 16-bit approximate adders}
   \adjustbox{max width=\columnwidth}{%
    \begin{tabular}{|c|c|>{\centering\arraybackslash}m{3cm}|>{\centering\arraybackslash}m{3cm}|>{\centering\arraybackslash}m{3cm}|}
    \hline
    16-bit Adder & Configuration & Delay Estimate using the proposed analytical Model ($nSec$) & Delay computed using Synopsys Design Compiler ($nSec$) & Accuracy of delay estimation model \\ \hline
    \multirow{15}{*}{HBAA} 
    & \{1,4,2,3\}\{4,0,3,1\} & 12.6 &	11.83&	93.49\% \\\cline{2-5}
      & \{4,2\}\{0,2\} & 30.24&	31.16&	97.05\%\\\cline{2-5}
      &\{2,2,2,1\}\{0,0,0,1\} & 35.2&	38.41&	91.64\%\\\cline{2-5}
        &\{6\}\{3\} & 30.24&	31.05&	97.39\%\\\cline{2-5}
        &\{2,1,2,1,0,2,1,1\}\{1,1,2,1,2,1,2,2\}& 7.56&	7.93&	95.33\%\\\cline{2-5}
        &\{3,2,3\}\{2,2,1\} & 12.6&	13.94&	90.39\%\\\cline{2-5}
        &\{4,4\}\{0,2\} & 30.24&	31.41&	96.28\%\\\cline{2-5}
        &\{5,2\}\{4,4\} & 10.7&	11.34&	94.36\%\\\cline{2-5}
        &\{2,1,2,1,0,2\}\{0,0,0,1,2,1\} & 20.16&	18.57&	91.44\%\\\cline{2-5}
        &\{2,2,2,2,1,2,2\}\{1,0,2,1,2,2,1\}	& 12.6&	11.52&	90.63\%\\\cline{2-5}
        &\{2,3\}\{0,4\} & 35.28&	34.18&	96.78\%\\\cline{2-5}
        &\{1,1,2,0,0,2\}\{0,1,1,1,2,0\}	&16.38&	16.94&	96.69\%\\\cline{2-5}
        &\{4,4\}\{2,3\}	&32.76&	32.89&	99.60\%\\\cline{2-5}
        &\{2,1,2,1,0,2\}\{1,1,0,1,2,1\}& 20.16&	21.67&	93.03\%\\\cline{2-5}
        &\{2,1,2,1,2,2,1,1\}\{1,2,2,1,2,0,2,2\} & 7.56&	6.94&	91.07\%\\\hline

     \multirow{3}{*}{SARA}
     &SARA2&10.08&	12.79&	78.81\%\\\cline{2-5}
    &SARA4&17.64&	21.46&	82.20\%\\\cline{2-5}
    &SARA8&27.72&	30.32&	91.42\%\\\hline
    \multirow{3}{*}{GeAr}  
    & \{16,4,0\}&	12.6&	13.76&	91.57\%\\\cline{2-5}
    &\{16,8,0\}&	22.68&	23.21&	97.72\%\\\cline{2-5}
    &\{16,6,4\}&	27.72&	27.23&	98.20\%\\\hline   
    \multirow{3}{*}{BCSA} 
    &BCSA2&	12.6&	11.29&	88.40\%\\\cline{2-5}
	&BCSA4&	20.16&	19.76&	97.98\%\\\cline{2-5}
	&BCSA8&	25.2&	28.2&	89.36\%\\\hline
    \end{tabular}}%
  \label{tab:hardware_delay_16}%
\end{table}

\begin{table}[ht]
  \centering
  \caption{Accuracy of our proposed estimation model for computing the Power of 16-bit approximate adders}
   \adjustbox{max width=\columnwidth}{%
    \begin{tabular}{|c|c|>{\centering\arraybackslash}m{3cm}|>{\centering\arraybackslash}m{3cm}|>{\centering\arraybackslash}m{3cm}|}
    \hline
    16-bit Adder & Configuration & Power Estimate using the proposed analytical Model ($\mu W$) & Power computed using Synopsys PrimeTime ($\mu W$) & Accuracy of power estimation model \\ \hline
    \multirow{15}{*}{HBAA}      
& \{1,4,2,3\}\{4,0,3,1\}&1224.96&	1468.34&	83.42\%\\\cline{2-5}
& \{4,2\}\{0,2\} &4085.9&	4275.3&	95.57\%\\\cline{2-5}
 &\{2,2,2,1\}\{0,0,0,1\}&4440.4&	3881.96&	85.61\%\\\cline{2-5}
&\{6\}\{3\}&4219.5&	4267.28&	98.88\%\\\cline{2-5}
 &\{2,1,2,1,0,2,1,1\}\{1,1,2,1,2,1,2,2\}&1218&	1127.67&	91.99\%\\\cline{2-5}
&\{3,2,3\}\{2,2,1\} &1476.52&	1756.34&	84.07\%\\\cline{2-5}
&\{4,4\}\{0,2\}&3784.5&	4473.2&	84.60\%\\\cline{2-5}
	 &\{5,2\}\{4,4\}&1552.52&	1694.72&	91.61\%\\\cline{2-5}
 &\{2,1,2,1,0,2\}\{0,0,0,1,2,1\}&2366.4	&2542.3	&93.08\%\\\cline{2-5}
	&\{2,2,2,2,1,2,2\}\{1,0,2,1,2,2,1\}&1684.32&	1765.43	&95.41\%\\\cline{2-5}
	&\{2,3\}\{0,4\}&5348.76	&5423.7	&98.62\%\\\cline{2-5}
&\{1,1,2,0,0,2\}\{0,1,1,1,2,0\}&2338.56&	2673.21&	87.48\%\\\cline{2-5}
 &\{4,4\}\{2,3\}& 4879.21	&5582.3	& 87.40\%\\\cline{2-5}
&\{2,1,2,1,0,2\}\{1,1,0,1,2,1\}&2721.36&	2957.42&	92.02\%\\\cline{2-5}
 &\{2,1,2,1,2,2,1,1\}\{1,2,2,1,2,0,2,2\}&1169.28&	1069.67	&90.69\%\\ \hline
     \multirow{3}{*}{SARA}
      &SARA2&2630.88&	3251.2&	80.92\%\\\cline{2-5}
 &SARA4&4750.2&	4896.3&	97.02\%\\\cline{2-5}
 &SARA8&6563.28&	5649.4&	83.82\%\\\hline 
    \multirow{3}{*}{GeAr}  
    & \{16,4,0\}&4746.72&	4237.9&	87.99\%\\\cline{2-5}
&\{16,8,0\}&5702.85&	5993.8&	95.15\%\\\cline{2-5}
 &\{16,6,4\}&7203.6&	7290.1&	98.81\%\\\hline 
    \multirow{3}{*}{BCSA} 
    &BCSA2&3970.68&	4164.2&	95.35\%\\\cline{2-5}
&BCSA4&4584.9&	4235.7&	91.76\%\\\cline{2-5}
&BCSA8&5517.54&	4988.2&	89.39\%\\\hline
    \end{tabular}}%
  \label{tab:hardware_power_16}%
\end{table}
\begin{table}[ht]
  \centering
  \caption{Accuracy of our proposed estimation model for computing the Area of 32-bit approximate adders}
   \adjustbox{max width=\columnwidth}{%
    \begin{tabular}{|c|c|>{\centering\arraybackslash}m{3cm}|>{\centering\arraybackslash}m{3cm}|>{\centering\arraybackslash}m{3cm}|}
    \hline
    32-bit Adder Type&	Configuration&	Area Estimate using the proposed analytical Model  ($\mu m^2$)&	Area computed using Synopsys Design Compiler  ($\mu m^2$)& 	Accuracy of area estimation model \\ \hline
     \multirow{3}{*}{HBAA}
     &\{2,2,2,2,2,2,1\}\{0,0,0,0,0,0,0,1\}	&27.23	&25.02	&91.17\%\\\cline{2-5}
     &\{4,4,4,2\}\{0,0,0,2\}&	30.19&	28.8&	95.17\%\\\cline{2-5}
     &\{8,4\}\{0,4\}&	31.49&	30.43&	96.52\%\\\hline
    \multirow{3}{*}{SARA} 
    &SARA2&	71.73&	65.86&	91.09\%\\\cline{2-5}
    &SARA4&	65.71&	62.1&	94.19\%\\\cline{2-5}
    &SARA8&	49.65&	50.81&	97.72\%\\\hline
    \multirow{3}{*}{GeAr}  
    &\{32,2,2\}&	40.54&	38.1&	93.60\%\\\cline{2-5}
    &\{32,4,4\}&	56.27&	53.34&	94.51\%\\\cline{2-5}
    &\{32,8,2\}&	69.86&	66.05&	94.23\%\\\hline
    \multirow{3}{*}{BCSA}
      &BCSA2&	41.23&	38.69&	93.43\%\\\cline{2-5}
    &BCSA4&	44.16&	45.52&	97.01\%\\\cline{2-5}
    &BCSA8&	53.47&	56.9&	93.97\%\\\hline
    \end{tabular}}%
  \label{tab:hardware_area32}%
\end{table}

\begin{table}[ht]
  \centering
  \caption{Accuracy of our proposed estimation model for computing the Delay of 32-bit approximate adders}
   \adjustbox{max width=\columnwidth}{%
    \begin{tabular}{|c|c|>{\centering\arraybackslash}m{3cm}|>{\centering\arraybackslash}m{3cm}|>{\centering\arraybackslash}m{3cm}|}
    \hline
    32-bit Adder Type&	Configuration&	Delay Estimate using the proposed analytical Model  ($nSec$)&	Delay computed using Synopsys Design Compiler  ($nSec$)& 	Accuracy of delay estimation model \\ \hline
     \multirow{3}{*}{HBAA}
     &\{2,2,2,2,2,2,1\}\{0,0,0,0,0,0,0,1\}&44.56&	48.67&	91.56\%\\\cline{2-5}
    &\{4,4,4,2\}\{0,0,0,2\}&60.28&	61.87&	97.43\%\\\cline{2-5}
    &\{8,4\}\{0,4\}&65.38&	63.27&	96.67\%\\\hline
    \multirow{3}{*}{SARA} 
    &SARA2&39.63&	41.15&	96.31\%\\\cline{2-5}
    &SARA4&46.27&	49.65&	93.19\%\\\cline{2-5}
    &SARA8&65.49&	70.97&	92.28\%\\\hline
    \multirow{3}{*}{GeAr}  
    &\{32,2,2\}&13.65&	14.71&	92.79\%\\\cline{2-5}
&\{32,4,4\}&18.27&	20.85&	87.63\%\\\cline{2-5}
&\{32,8,2\}&23.94&	25.63&	93.41\%\\\hline
    \multirow{3}{*}{BCSA}
      &BCSA2&20.21&	18.45&	90.46\%\\\cline{2-5}
&BCSA4&26.84&	23.38&	85.20\%\\\cline{2-5}
&BCSA8&24.76&	27.29&	90.73\%\\\hline
    \end{tabular}}%
  \label{tab:hardware_delay32}%
\end{table}

\begin{table}[ht]
  \centering
  \caption{Accuracy of our proposed estimation model for computing the Power of 32-bit approximate adders}
   \adjustbox{max width=\columnwidth}{%
    \begin{tabular}{|c|c|>{\centering\arraybackslash}m{3cm}|>{\centering\arraybackslash}m{3cm}|>{\centering\arraybackslash}m{3cm}|}
    \hline
    32-bit Adder Type&	Configuration&	Power Estimate using the proposed analytical Model  ($\mu W$)&	Area computed using Synopsys PrimeTime  ($\mu W$)& 	Accuracy of power estimation model \\ \hline
     \multirow{3}{*}{HBAA}
     &\{2,2,2,2,2,2,1\}\{0,0,0,0,0,0,0,1\}&11568.4&	12675.7&	91.26\%\\\cline{2-5}
    &\{4,4,4,2\}\{0,0,0,2\}&18326.15&	19876.5&	92.20\%\\\cline{2-5}
    &\{8,4\}\{0,4\}&20699.43&	22316.9&	92.75\%\\\hline
    \multirow{3}{*}{SARA} 
    &SARA2&28931.34&	31427.6&	92.06\%\\\cline{2-5}
    &SARA4&30807.04&	34697.2&	88.79\%\\\cline{2-5}
    &SARA8&32690.47&	38246.8&	85.47\%\\\hline
    \multirow{3}{*}{GeAr}  
    &\{32,2,2\}&7984.2&	8561.2&	93.26\%\\\cline{2-5}
    &\{32,4,4\}&10840.01&	11864.1&	91.37\%\\\cline{2-5}
    &\{32,8,2\}&16249.15&	17649.2&	92.07\%\\\hline
    \multirow{3}{*}{BCSA}
     &BCSA2&9138.335&	10264.3&	89.03\%\\\cline{2-5}
    &BCSA4&12240.13&	13468.4&	90.88\%\\\cline{2-5}
    &BCSA8&13723.6&	15237.1&	90.07\%\\\hline
    \end{tabular}}%
  \label{tab:hardware_power32}%
\end{table}

\subsection{Comparison of HBAA with State-of-the-art Approximate Adders}

In this section, we compare the HBAA with state-of-the-art approximate adders, i.e., GeAr, SARA, BCSA, $QuAd_o$ and conventional LPAAs~\cite{6387646}\cite{7459392} configurations shown in Table~\ref{tab:tabtruth}. 
For the comparison, we computed all the hardware metrics, i.e., area, delay, and power, of all HBAA and other state-of-the-art approximate adder configurations using our proposed hardware estimation models. 
For error metrics such as MED, we used our proposed analytical model for all HBAA configurations and Monte Carlo (MC) simulations for all GeAr, SARA, BCSA, $QuAd_o$, and conventional LPAA~\cite{6387646}\cite{7459392} configurations. 
Note that we used exhaustive simulations with $2^{16}$ input combinations for 8-bit approximate adders. 

\begin{figure*}[ht]
    \subfloat[Delay vs. MED]{
        \includegraphics[width=0.5\columnwidth]{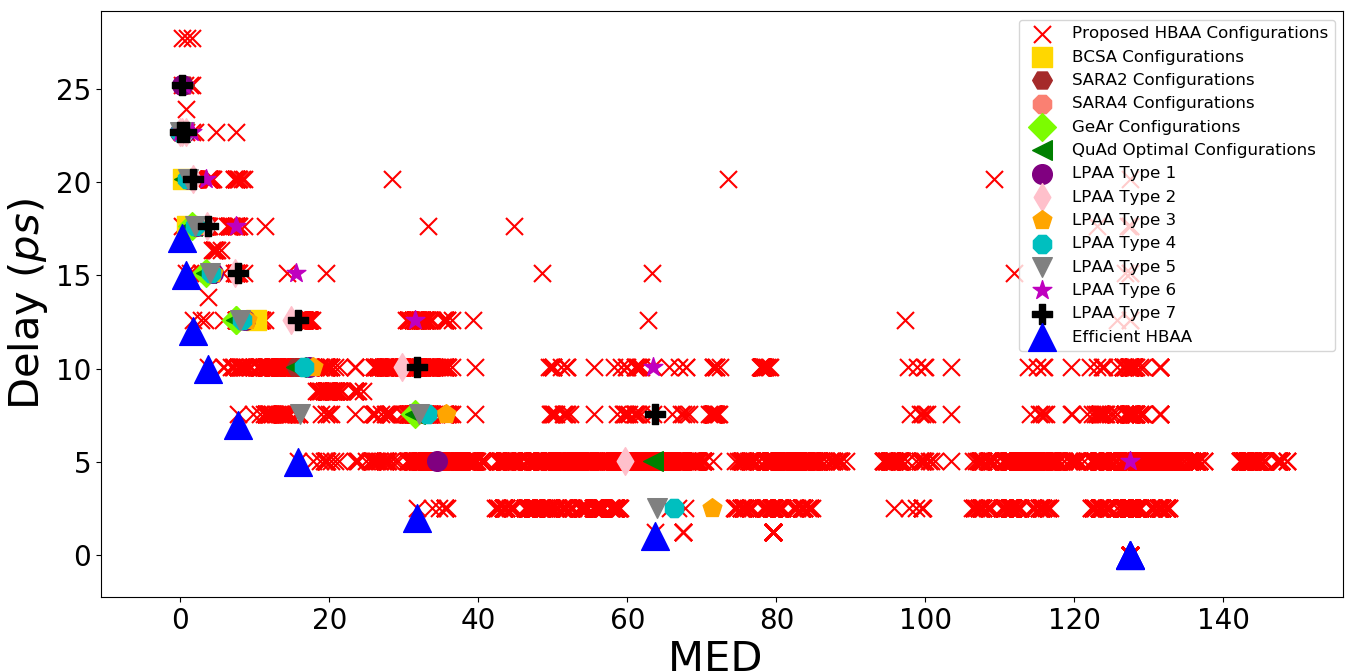}
        
        \label{figure_optimal delay_zoom}
    }
    \subfloat[Area vs. MED]{
         \includegraphics[width=0.5\columnwidth]{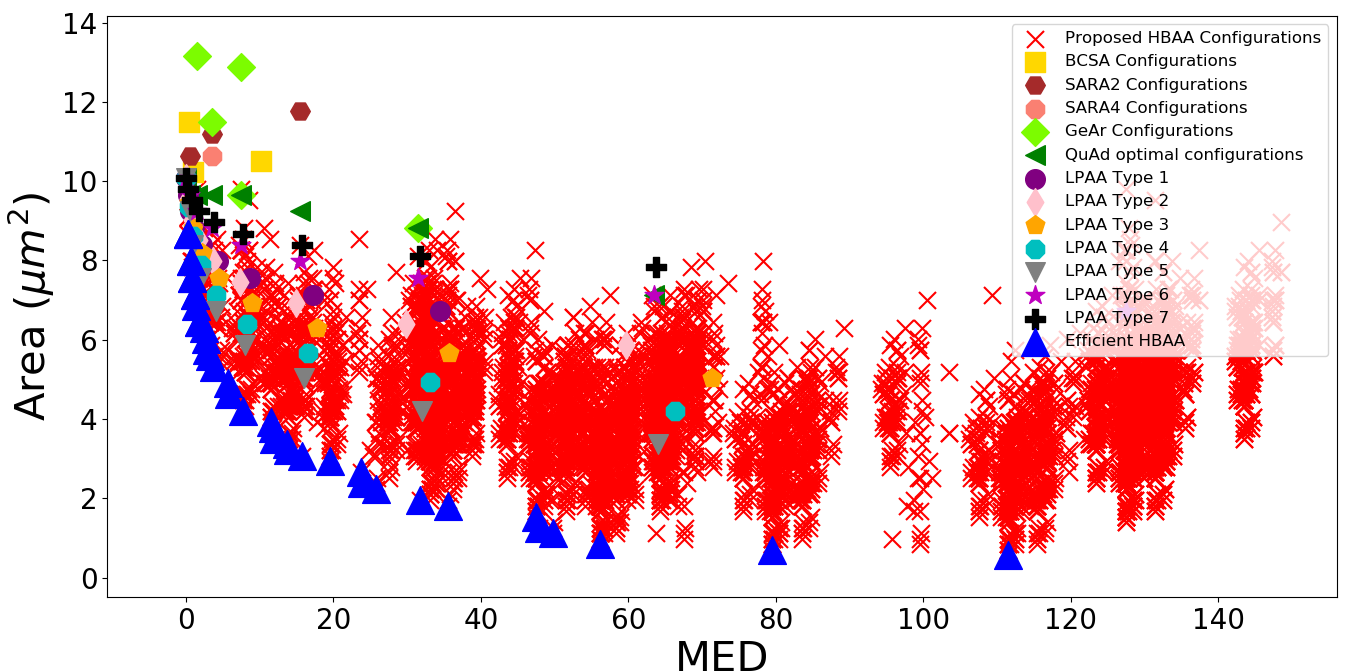}
        
         \label{figure_optimal area}
    }\hfill
    \subfloat[Power vs. MED]{
         \centering
         \includegraphics[width=0.5\columnwidth]{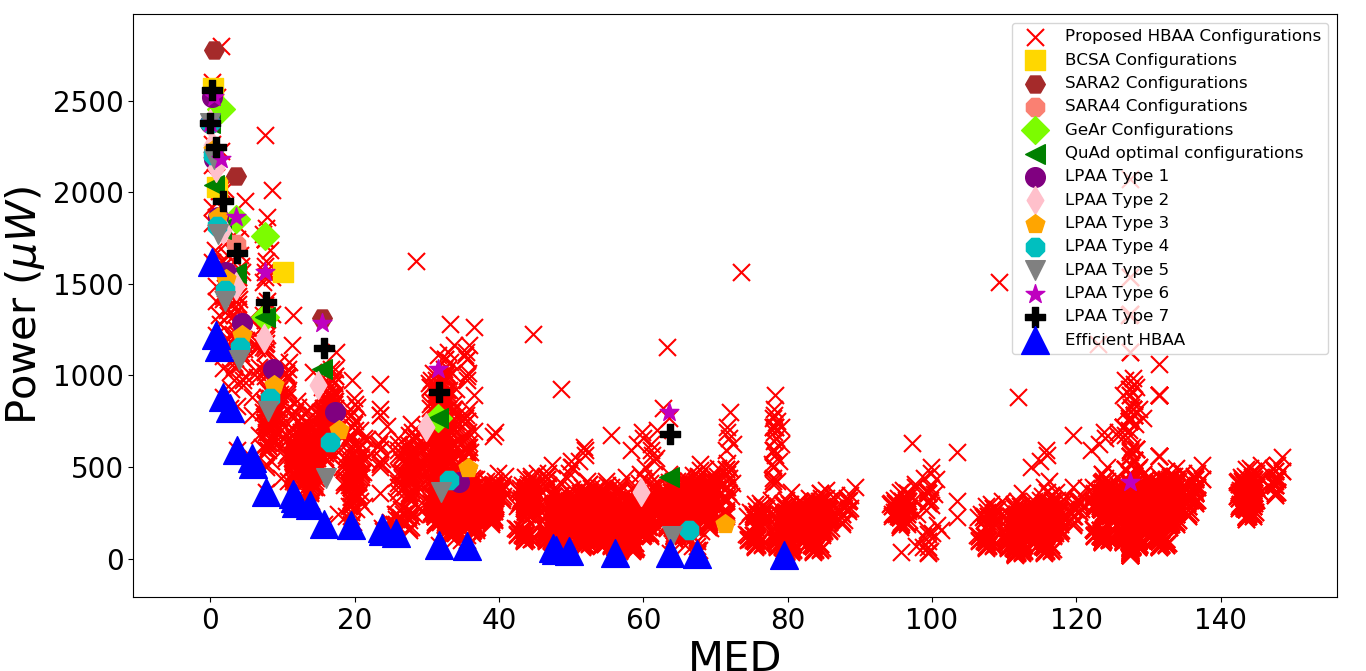}
         \label{figure_optimal power}
    }\hfil
      \caption{Design space for 8-bit approximate adder based on HBAA, GeAr, SARA, BCSA, $QuAd_o$ and conventional LPAA adder designs. The Pareto-optimal HBAA configurations are marked using `$\begingroup\color{blue}\blacktriangle\endgroup$' symbol.}
      \label{fig:design_space_8-bit}
\end{figure*}

\begin{figure*}[ht]
    \subfloat[Delay vs. NMED]{
        \includegraphics[width=0.5\columnwidth]{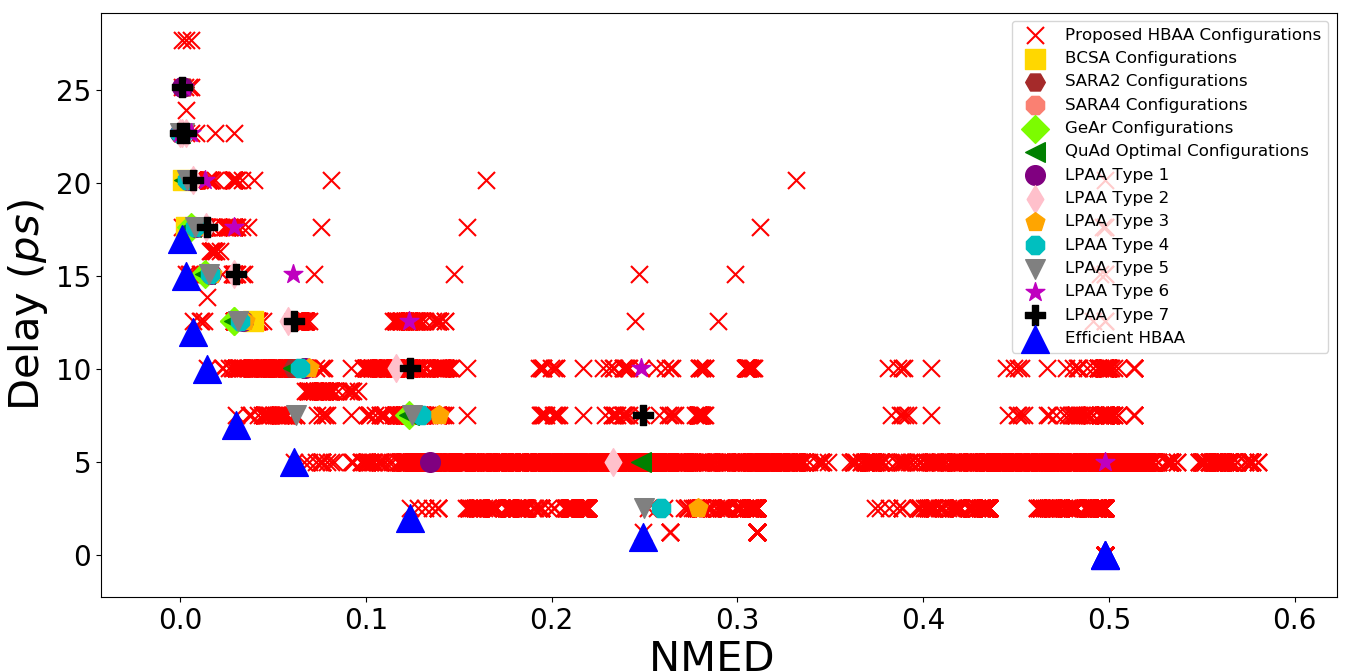}
        
        \label{figure_NMED}
    }
    \subfloat[Delay vs. ER]{
         \includegraphics[width=0.5\columnwidth]{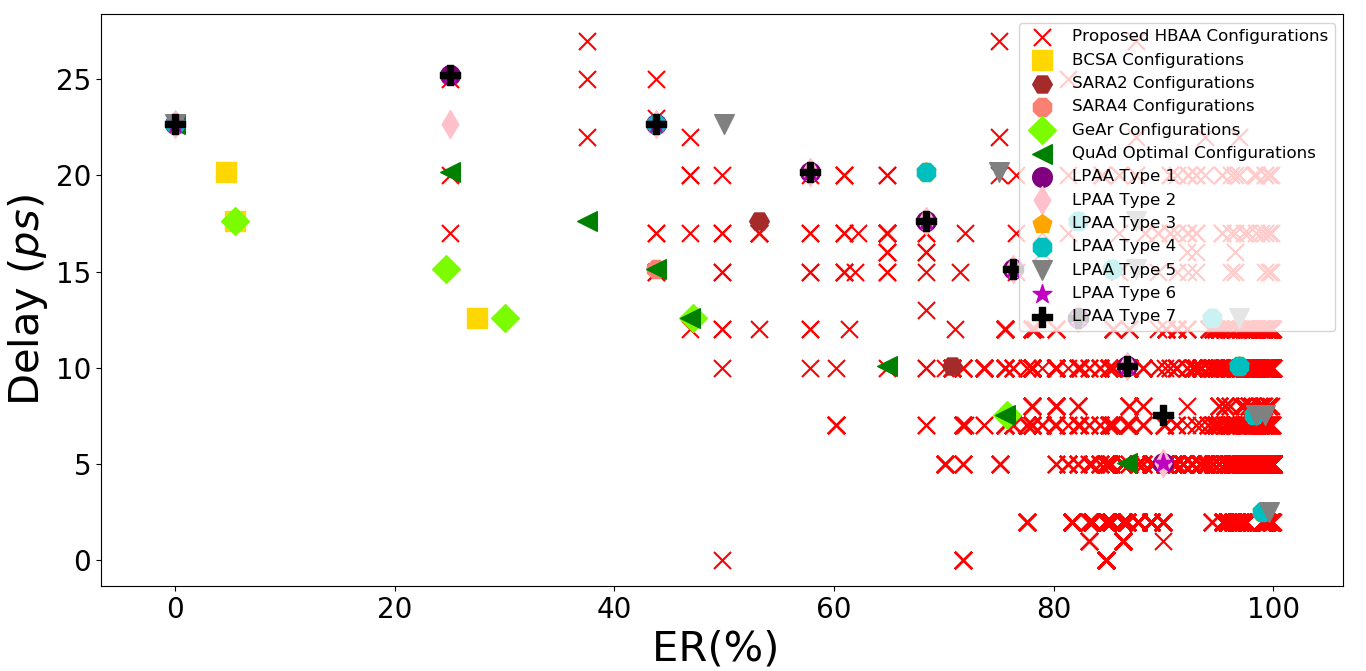}
        
         \label{figure_ER}
    }\hfill
      \caption{Design space for 8-bit approximate adder based on HBAA, GeAr, SARA, BCSA, $QuAd_o$ and conventional LPAA adder designs. The Pareto-optimal HBAA configurations are marked using `$\begingroup\color{blue}\blacktriangle\endgroup$' symbol.}
      \label{fig:design_space_8-bit_NMED_ER}
\end{figure*}

\begin{figure*}[ht]

    \subfloat[Delay vs. MED]{
        \includegraphics[width=0.5\columnwidth]{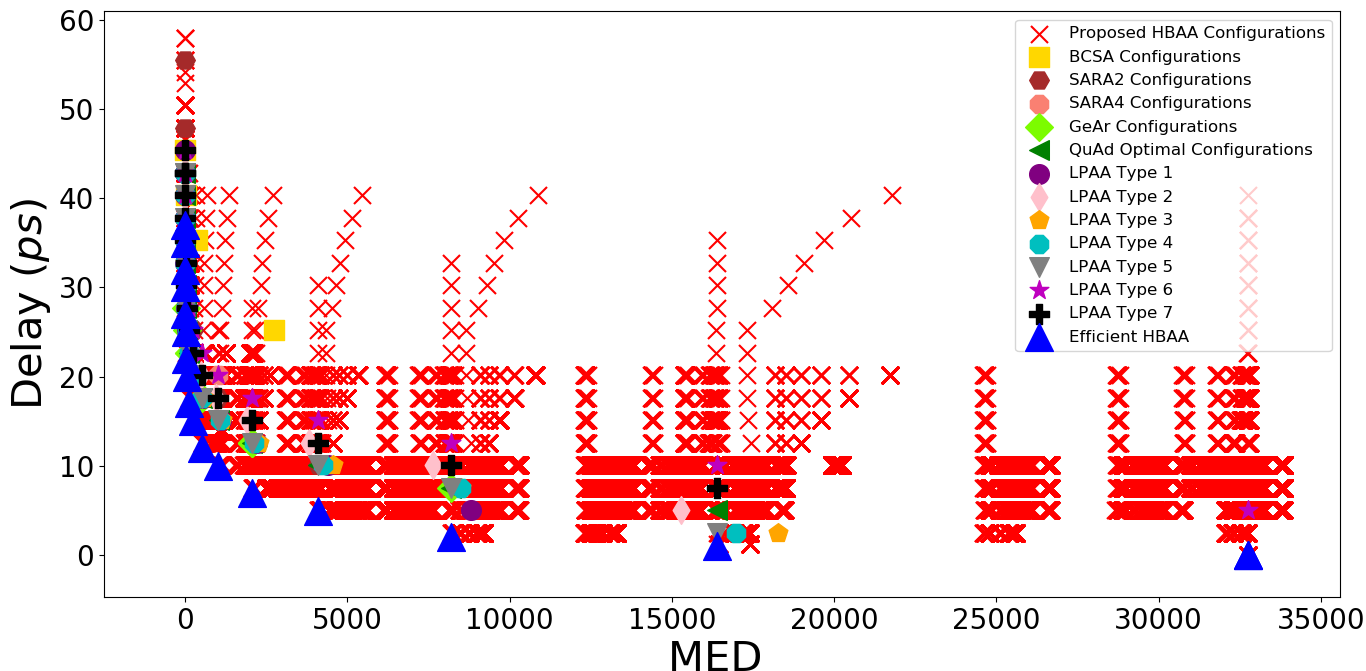}
        
        \label{figure_optimal delay_zoom_16}
    }
    \subfloat[Area vs. MED]{
         \includegraphics[width=0.5\columnwidth]{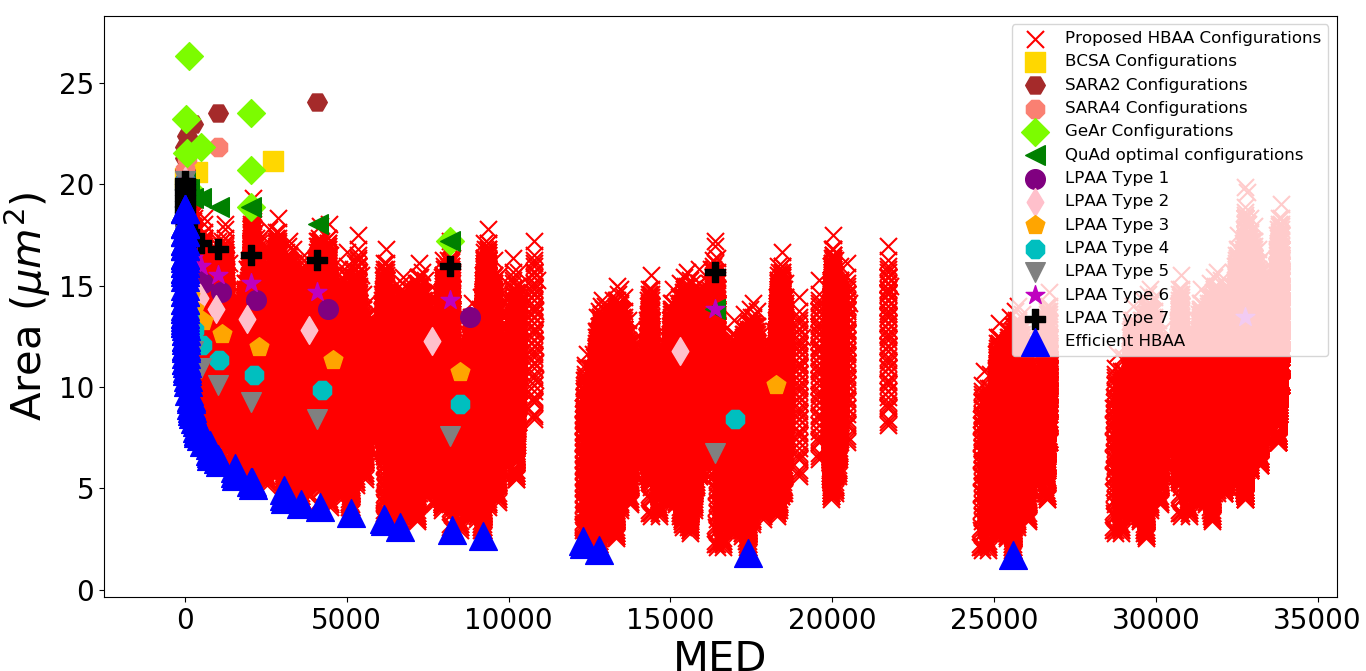}
        
         \label{figure_optimal area_16}
    }\hfil
    \subfloat[Power vs. MED]{
         \centering
         \includegraphics[width=0.5\columnwidth]{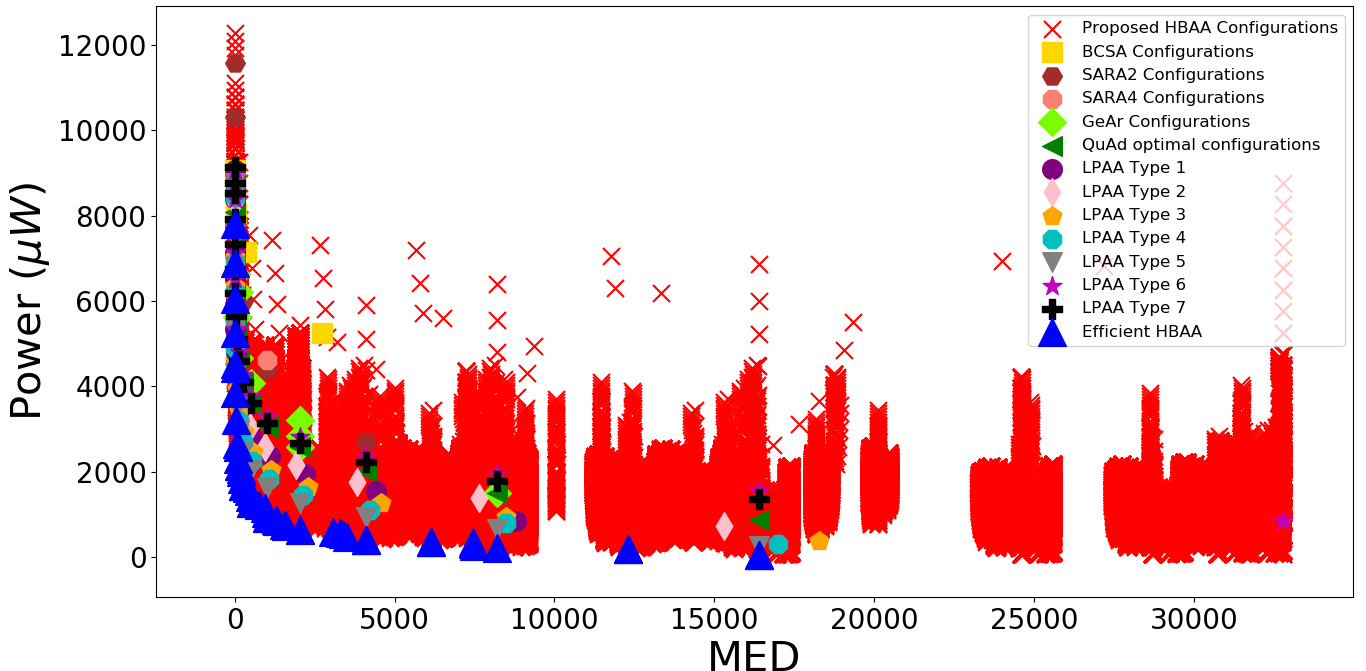}
         \label{figure_optimal power_16}
    }\hfil
      \caption{Design space for 16-bit approximate adder based on HBAA, GeAr, SARA, BCSA, $QuAd_o$ and conventional LPAA adder designs. The Pareto-optimal HBAA configurations are marked using `$\begingroup\color{blue}\blacktriangle\endgroup$' symbol.}
      \label{fig:design_space_16-bit}
\end{figure*}

Different approximate adders offer different accuracy-efficiency trade-offs. 
Based on the user requirements, a design space exploration is usually required to find optimal configurations that offer the best output quality while meeting the user-defined resource constraints. 
Figures~\ref{fig:design_space_8-bit},~\ref{fig:design_space_8-bit_NMED_ER}, and~\ref{fig:design_space_16-bit} show the design points for 8-bit and 16-bit approximate adders composed of equal-sized sub-adders. 
It can be observed from the figures that in all cases, i.e., area vs. MED, delay vs. MED, power vs. MED, and delay vs. NMED, HBAA configurations offer the best quality-efficiency trade-off compared to GeAr, SARA, BCSA, $QuAd_o$ and conventional LPAA configurations. 
However, in the case of delay vs. ER, some of the state-of-the-art approximate adder configurations offer better results compared to HBAA. 
Note that, in the most of the cases, metrics that are a measure of error magnitude are considered more important than simple error rate. 
\textit{Thus, from this analysis, it can be concluded that HBAA introduces additional configurations in the approximate adder design space that can offer better results compared to state-of-the-art approximate adder designs.} 

{}

{}

{}

\subsection{Execution Time for Design Space Exploration using the Proposed Analytical Models}

\begin{figure}
    \centering
    \includegraphics[scale=0.6]{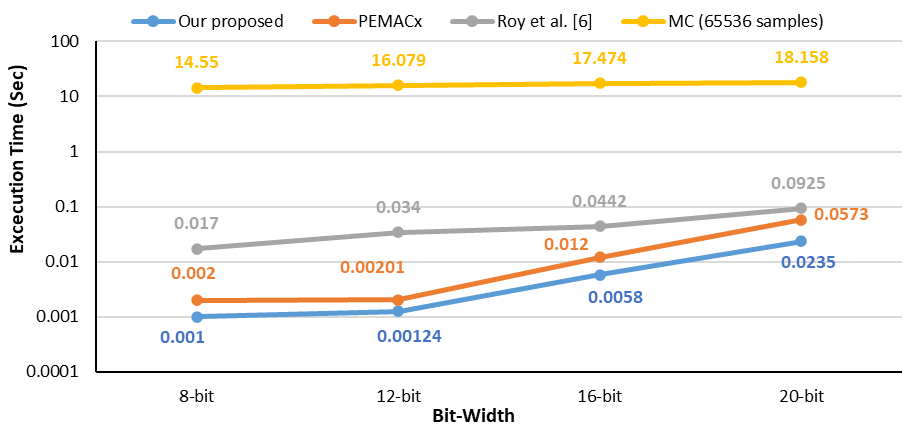}
    \caption{Execution Time comparison of the MED computation algorithms}
    \label{execute_run}
\end{figure}

We have also compared the execution time of the proposed analytical model for computing MED with Monte Carlo (MC) simulations and state-of-the-art error estimation methods such as PEMACx~\cite{hanif2020pemacx} and Roy~et~al.~\cite{8994186}. 
For Monte Carlo simulations in this section, we used $2^{16}$ randomly selected input combinations. 
The execution time of all the above-mentioned error estimation methods for different adder bit-widths (i.e., 8-bit to 20-bit) is shown in Figure~\ref{execute_run}.
  {}
It can be observed from the figure that our proposed analytical model is faster than the other existing analytical models for computing the error estimates. 
For example, for 16-bit HBAA, our proposed model is about 6~times and 21~times faster than PEMACx~\cite{hanif2020pemacx} and Roy et al.~\cite{8994186}, respectively.

\begin{figure}[ht]
    \centering
    \includegraphics[width=0.5\columnwidth]{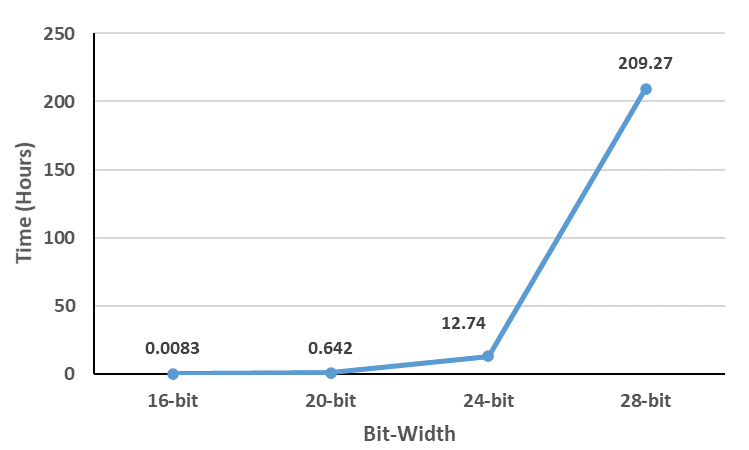}
    \caption{Executing time to find the efficient configuration of HBAA}
    \label{timerun}
\end{figure}

The overall design space exploration time to find the best HBAA configuration for a given set of user-defined resource constraints depends on the bit-width of the adder. 
Figure~\ref{timerun} presents the time required to find the best HBAA configuration at different bit-widths. 
The figure shows that with the increase in bit-width the execution time increases significantly. 
This is mainly because, as the adder size increases, the number of sub-adders increases and the number of combinations of different sub-adder configurations increases exponentially. 
Therefore, the speed of our proposed algorithm reduces significantly due to the exponential increase in the number of computations and memory size. 
To understand this exponential increase in the number of sub-adder combinations, consider an $N$-bit HBAA constructed using $H$-bit sub-adders. Given the architecture of HBAA, each approximate sub-adder can have $C_H=(H+1)\times(H+1) - 1$ different configurations. Moreover, given that an $N$-bit HBAA has in total $k=[N/H]$ sub-adders and if $i^{th}$ sub-adder is approximate then all the less significant $i-1$ sub-adders should also be approximate, we get total approximate configurations for an $N$-bit HBAA with $H$-bit sub-adders equals $\sum_{i=1}^{k}C^i$. Thus, it can be said that (in general) the total number of configurations of HBAA increases exponentially with the increase in the number of sub-adders and the size of the adder. 

 {}

 \section{Conclusion}
 \label{Sec:Conclusion}

In this paper, we present a new class of energy-efficient approximate adders, namely Heterogeneous Block-based Approximate Adders (HBAA), and propose a generic configurable adder model that can be configured to represent a particular HBAA configuration. 
An HBAA, in general, is composed of heterogeneous sub-adder blocks of equal length, where each sub-adder can be an accurate or approximate sub-adder and have a different configuration. 
The sub-adders are mainly approximated through inexact logic and carry truncation. 
To enable efficient design space exploration based on user-defined constraints, we proposed an analytical model to efficiently compute the PMF of error and other error metrics, e.g., MED, ER, and NMED of HBAAs.
Moreover, we present hardware estimation models for the computing area, delay, and power of HBAAs. 
Our results showed that compared to the design space of existing approximate adders, HBAA provides additional design points that offer a better quality-efficiency trade-off.

\bibliographystyle{ACM-Reference-Format}
\bibliography{Main}
\vspace{5mm}

\end{document}